\documentclass[12pt]{article}
\usepackage[authoryear]{natbib}
\newcommand{\blind}{0}
\usepackage[footnotesize]{caption} 
\usepackage{amssymb}
\usepackage{epsfig}
\usepackage{amsmath}
\usepackage{graphicx}
\usepackage{graphics}
\usepackage{lscape}
\usepackage{float}
\usepackage{subfigure}
\usepackage{multirow}
\usepackage[toc,page]{appendix}
\usepackage{setspace}
\usepackage{color}

\usepackage{soul,xcolor}
\setstcolor{red}

\newcommand{\bm}[1]{\mbox{\boldmath $#1$}}
\newcommand{\be}{\begin{equation}}
\newcommand{\ee}{\end{equation}}

\newcommand{\nd}{\noindent}

\newcommand{\mc}[1]{\mathcal{#1}}

\newcommand{\RN}[1]{%
  \textup{\uppercase\expandafter{\romannumeral#1}}%
  }
\newcommand{\la}{\left \langle}
\newcommand{\ra}{\right \rangle}

\graphicspath{{Figures/}}
\begin{document}
\title{\vspace{-1.5cm} An Intrinsic Geometrical Approach for Statistical Process Control of Surface and Manifold Data}
\if0\blind
{
\author{
{\small Xueqi Zhao}\\
{\small Enrique del Castillo\footnote{Corresponding author. Dr. Castillo is Distinguished Professor of Industrial \& Manufacturing Engineering and Professor of Statistics. e-mail: exd13@psu.edu}}\\  
{\small Engineering Statistics and Machine Learning Laboratory}\\
{\small Department of Industrial and Manufacturing Engineering and Dept. of Statistics}\\
{\small The Pennsylvania State University, University Park, PA 16802, USA}}\vspace{0.3cm}

\date{\footnotesize \today\vspace{-1.5cm}}
}\fi
\maketitle
\begin{abstract}
We present a new method for statistical process control (SPC) of a discrete part manufacturing system based on intrinsic geometrical properties of the parts estimated from 3-dimensional (3D) sensor data. An intrinsic method has the computational advantage of avoiding the difficult part registration problem, necessary in previous SPC approaches of 3D geometrical data, but inadequate if non-contact sensors are used. The approach estimates the spectrum of the Laplace-Beltrami (LB) operator of the scanned parts and uses a multivariate nonparametric control chart for on-line process control. Our proposal brings SPC closer to computer vision and computer graphics methods aimed to detect large differences in shape (but not in size). However, the SPC problem differs in that small changes in either shape or size of the parts need to be detected, keeping a controllable false alarm rate and without completely filtering noise. An on-line or ``Phase II"  method and a scheme for starting up in the absence of prior data (``Phase I") are presented. Comparison with earlier approaches that require registration shows the LB spectrum method to be more sensitive to rapidly detect small changes in shape and size, including the practical case when the sequence of part datasets is in the form of large, unequal size meshes.  A post-alarm diagnostic method to investigate the location of defects on the surface of a part is also presented. While we focus in this paper on surface (triangulation) data, the methods can also be applied to point cloud and voxel metrology data.
\end{abstract}

\newpage
 \setcounter{page}{2}
\baselineskip=16pt

\section{INTRODUCTION}

Widespread use of modern sensors in engineering and industry in recent times has resulted not only in bigger datasets but also in more complex datasets. Statistical Process Control (SPC) is an area that has witnessed increased sophistication in metrology accompanied by increased complexity in the resulting data sets related to production or manufacturing processes (\citealt{colosimo2014profile,Colosimo2018}). We consider the case non-contact sensors (or a combination of contact and non-contact sensors) collect geometric data formed by thousands of points which actually lie on what could be described as a 2-dimensional curved space or manifold, that is, the surface of the object or part produced. Due to manufacturing and measurement errors, the observed surface will deviate from the target or nominal surface, given by the design of the part, usually available in some Computer Aided Design (CAD) file. 

The purpose of this paper is to lay the foundations of a new methodology for SPC of discrete-part 3D geometrical data that can be in either of the form of a point cloud, a mesh,  {or,}  even voxel (volumetric) datasets, although we focus in this paper in the mesh case, the most common type of part data generated by non-contact scanners today. Prior approaches to this problem  utilize methods where the points scanned in each part need to be registered (or ``superimposed" one to one) and require that the exact same number of points in each part be scanned, tenable assumptions only when data are exclusively collected using contact sensors. Our main contributions are to provide a new SPC methodology for 3D geometrical data, based on intrinsic geometrical properties of datasets scanned from a sequence of parts, that: {\em i)} does {\em not} require registration of the points, meshes or voxel datasets scanned from each part, which is a computational expensive and non-convex problem for which no guarantee of global optimality can be given, and therefore, it is prone to error, and {\em ii)} does {\em not} require parts to have the exact same number of points. 

Our approach brings SPC of discrete-part manufacturing closer to the Computer Graphics/Vision fields. The approach we propose is based on techniques popular in computer graphics to characterize 3D objects, and these methods have also been used in Machine Learning of general manifolds of much higher dimension {than} the 2-manifolds (surfaces) we focus in this paper. In computer graphics and computer vision applications, however, the problem to solve is the identification of {\em large} differences, usually evident to the human eye, between the shapes of objects (frequently neglecting differences in size) for classification purposes or to match a query object. In computer graphics, one works with noise free meshes drawn by an artist or CAD engineer, while in computer vision noisy measurements are obtained but usually the noise is first filtered out.  Here, in contrast, we focus on detecting considerable smaller differences in either shape or size (sometimes not perceptible to the human eye) in the presence of measurement and manufacturing noise in a sequence of objects which the manufacturer is trying to produce consistently and with low variability, hence differences will tend to be rather small. Furthermore, in addition to detecting changes in the mean shape or size of an object, a noise increment could also be considered a process signature change to be detected, rather than a signal that must be filtered and ignored as in computer vision, where shape identification methods that are robust with respect to noise are typically sought.

Traditional treatment of 2 and 3-dimensional point cloud datasets in Statistics pertains to the field of Statistical Shape Analysis (SSA, \citealt{Kendall,DrydenMardia}; for applications in manufacturing see \citealt{EDCBianca}). In SSA, the $m$-point cloud data are represented by a configuration matrix ${\bm X} \in \mathbb{R}^{m \times n}$ with $n=2$ or 3. The {\em Shape} of an object is defined as the geometrical information in $\bm X$ that remains after discounting the effect of similarity transformations usually excepting reflections (translations, rotations and dilations). To make inferences on the shape of $N$ objects, the Generalized Procrustes Algorithm, {GPA,} is first applied. The GPA registers or superimposes all the $N$ objects by finding scaling factors $\beta_i \in \mathbb{R}$,
rotation matrices $\bm \Gamma_i \in SO(n)$ (the special orthogonal group, which excludes reflections and has determinant one) and $n-$dimensional
translation vectors $\bm \gamma_i, i=1,...,N$, such that they minimize
the sum of squared full procrustes distances between all pairs of configuration matrices ($d_F(\bm X_i, \bm X_j)$):
\begin{eqnarray}
\label{GPA}
G(\bm X_1, \bm X_2,..,\bm X_N) &=& \min_{\beta_i, {\scriptsize \bm \Gamma_i}, \bm \gamma_i} \frac{1}{N}
\sum_{i=1}^{N{-1}} \sum_{j=i+1}^N || \beta_i \bm X_i \bm \Gamma_i + \bm 1_m \bm \gamma_i' - (\beta_j \bm X_j \bm \Gamma_j + \bm 1_m \bm \gamma_j')||^2 \nonumber\\
&=& {\min_{\beta_i, {\scriptsize \bm \Gamma_i}, \bm \gamma_i}} \frac{1}{N} \sum_{i=1}^{N{-1}} \sum_{j=i+1}^N d^2_F(\bm X_i, \bm X_j) 
\end{eqnarray}
where $\bm 1_m$ is a vector of $m$ ones.  Constraints must be added to avoid the trivial solution where all parameters are zero \citep{DrydenMardia}. Note how two objects with different size{s} may still have the same shape, given that the effect of dilations (changes of scale) is usually filtered out in SSA. Neglecting differences in size is an aspect in common to shape classification in computer vision. Other shape analysis methods based on euclidean distances between the points \citep{Lele1993} require large distance matrices and will not be reviewed further here.

The problem of registering different three dimensional objects each with a large but {\em different} number of (non-corresponding) points has been known for a long time in the computer vision literature, where the Iterated Closest Point (ICP) algorithm (\citealt{Besl,Zhang}) is a standard. 
 Consider the configurations of two distinct {\em unlabeled} objects  $\bm X_q \in \mathbb{R}^{m_1 \times n} $ and $\bm X_p \in \mathbb{R}^{m_2 \times n}$ (with $n=3$), not necessarily having the same pose and assume $m_1 \leq m_2$. Let $\bm M \in \mathbb{R}^{m_1 \times m_2}$ with $M_{ij}=1$ if $\bm x_{q,i} \in \bm X_q$ is matched with point $\bm x_{p,j} \in \bm X_p$, and zero otherwise. The objects may be located and oriented differently in space, and hence the problem is not only to find the correspondences but also a rigid body transformation 
 $T(\bm x) = \bm \Gamma \bm x+ \bm \gamma$ that registers the two objects such that the following problem is solved: 
\be   \min_{\bm M, \bm \Gamma, \bm \gamma} L(\bm M, \bm \Gamma, \bm \gamma) = {\min_{\bm M, \bm \Gamma, \bm \gamma}}\sum_{i=1}^{m_1} \sum_{j=1}^{m_2} M_{ij} \cdot C(\bm \Gamma \bm x_{q,i}+ \bm \gamma, \bm x_{p,j}) 
\label{ICP}
\ee
subject to $ \sum_{j=1}^{m_2} M_{ij}=1, i=1,...,m_1$, and $M_{ij} = 0$ or 1,
where  {$C(\bm \Gamma \bm x_{q,i}+ \bm \gamma, \bm x_{p,j})$} is the cost of matching point {$\bm x_{q,i}$} to point {$\bm x_{p,j}$}. 
This is a hard non-linear discrete optimization problem. Existing heuristics differ by choosing different cost functions $C$. In the ICP method,  $C(\bm \Gamma \bm x_{q,i}+ \bm \gamma, \bm x_{p,j})= ||\bm \Gamma \bm x_{q,i}+ \bm \gamma - \bm x_{p,j}||$, the euclidean distance between $\bm x_{q,i}$ in $\bm X_q$ and {its} {\em closest point} $\bm x_{p,j}$ in $\bm X_p$ {after the transformation}.

\begin{figure}[h]
     \begin{center}
     \begin{tabular}{ll}
     \resizebox{4.5cm}{4cm}{\includegraphics{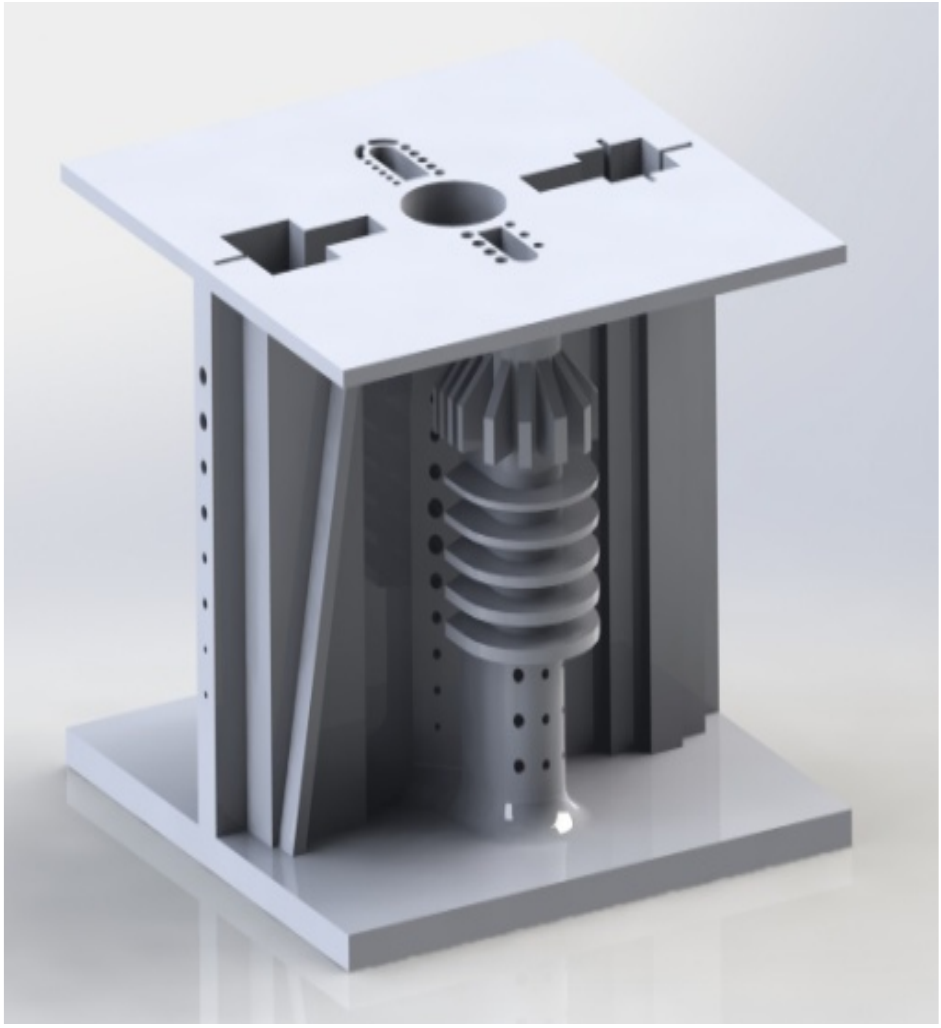}}&\quad \quad \quad
     \resizebox{4.5cm}{4cm}{\includegraphics{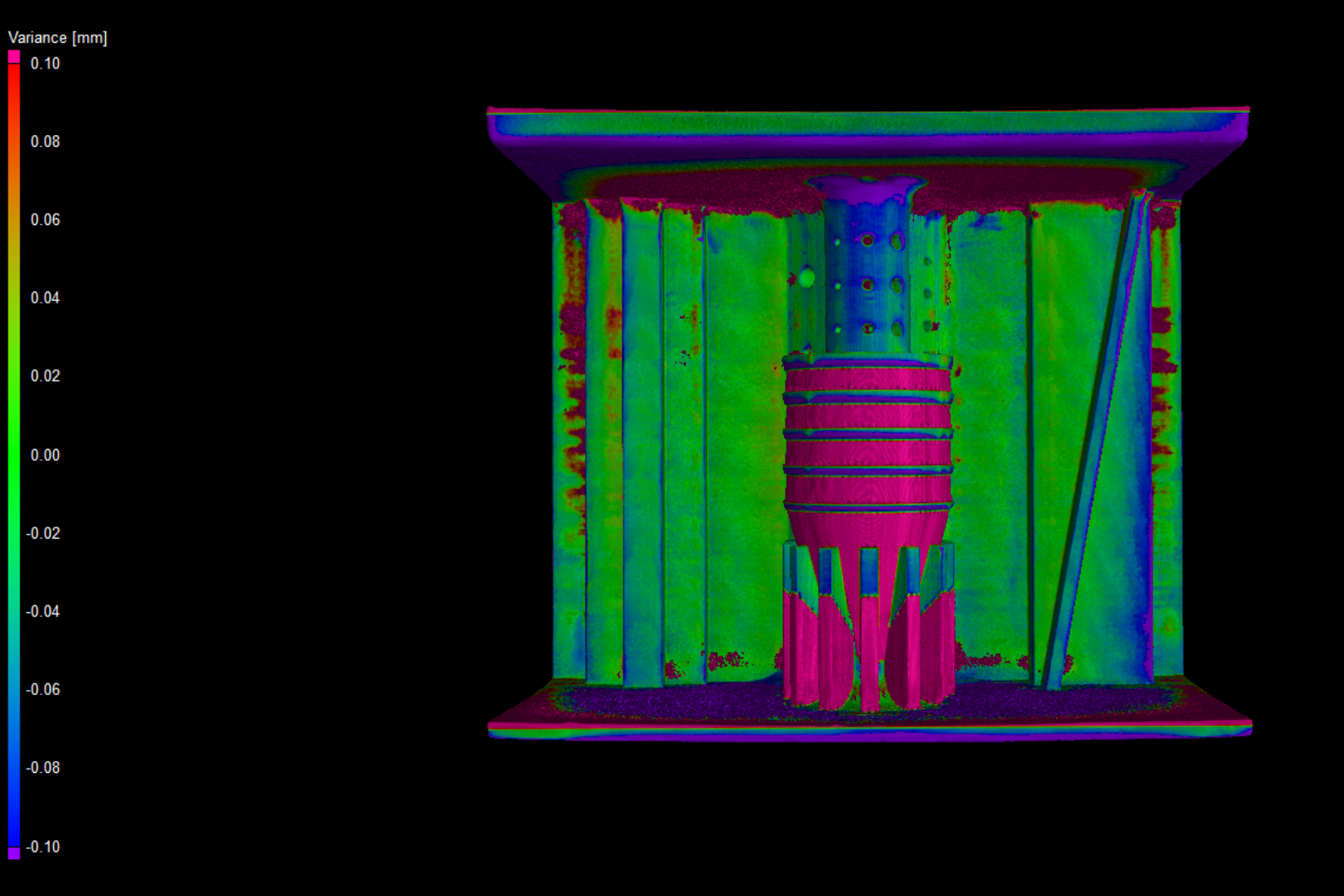}}
     \end{tabular}
      \end{center}
           \caption{Left: CAD design 3D rendering of a metal part. Right: CT image of the manufactured part with color contrast indicating differences in the $x$ dimension between the CAD model and the actual part. The CT software registers the CAD model and actual part using the ICP algorithm and color codes the deviations from nominal for visualization. SPC methods based on the deviations from nominal require computationally hard registration.
           }
           \label{fig2}
\end{figure}

Commercial CAD and inspection software use variants of the ICP method to align the cloud points, meshes or voxel datasets of each scanned object and that of the CAD file, in order to determine regions in the manufactured part that differ from nominal. For instance, to do this in the mesh data case, a CAD model, usually in the form of NURBS curves, is sampled to form a mesh or triangulation, after which the alignment of the CAD and scanned part triangulations can be performed. Figure \ref{fig2} shows an instance of a metal part CAD model and {a} color-coded comparison between the CAD model and the manufactured part (this is actually voxel data, but the registration problem is essentially identical). 
A first approach we will consider as a benchmark for Statistical Process Control of part surface  data is based on monitoring the deviations from nominal shown in Figure \ref{fig2} after applying the ICP method, similarly to a  ``DNOM" (deviation from nominal) control chart (see e.g.,  \citealt{Farnum}). The deviations from nominal are vectors, and either their norm or their individual components could be used for SPC. The optimal value of the ICP statistic (\ref{ICP}), {$L^*(\bm M, \bm \Gamma, \bm \gamma)$} between CAD file and each scanned part could be monitored on-line, for instance, to provide a ``generic" SPC mechanism against a wide variety of unanticipated out of control states in the geometry of a part, in a similar sense to what \cite{BoxRamirez} thought of univariate Shewhart charts, with other SPC or diagnostic mechanisms added to detect more specific, or more localized defects on the part. As far as we know, this simple strategy has not been proposed before in the literature, so we contrast it with the new intrinsic differential-geometrical methods that conform our main contribution and with some earlier SPC approaches that also require registration.

\begin{figure}[H]
\begin{center}
\includegraphics[scale=0.45]{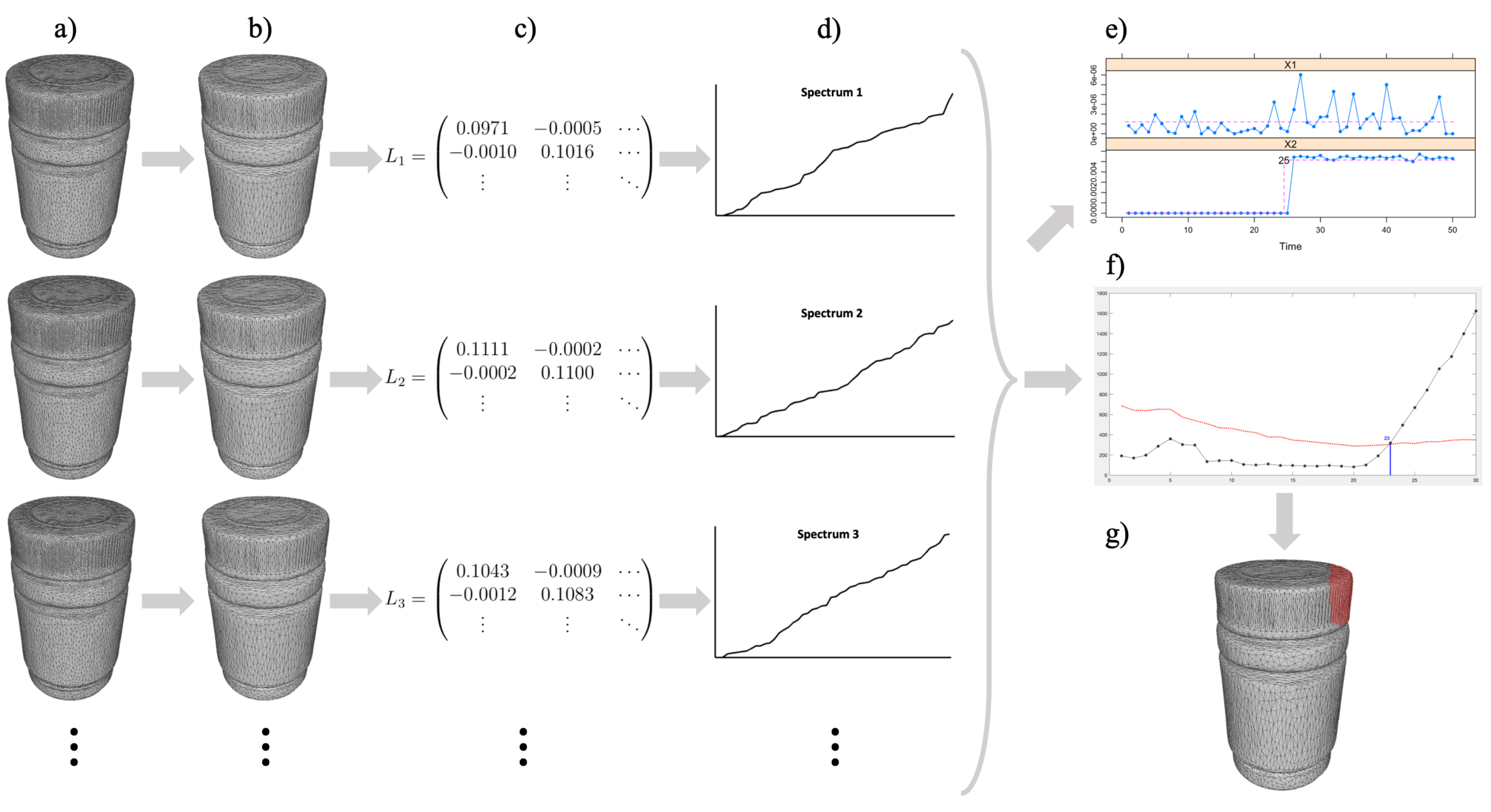}
\end{center}
\caption{The ``pipeline" of methods and algorithmic steps (overall proposed methodology). a) Objects are scanned and their meshes obtained; b) preprocessing of the meshes (optional); c) estimation of the Laplace-Beltrami (LB) operator of the surface $L_{\cal K}^t$ from each mesh; d) computation of the lower spectrum of each estimated LB operator; e) use \cite{capizzi2017phase} multivariate  permutation-based method for Phase I SPC applied to the lower LB spectra; f) once enough ($m_0$) in-control (IC) parts are collected, start on-line monitoring (Phase II) using \cite{Chen2016} distribution-free DFEWMA chart based on the lower LB spectra;  g) if the DFEWMA chart triggers an out of control alarm, perform diagnostics, including locating the change point part and locating the defects on the part using an Iterated Closest Point (ICP) diagnostic.}
\label{fig:pipeline}
\end{figure}

In this paper, we follow the traditional SPC paradigm with modern statistical tools. The main goal is detection of significant part to part differences with respect to historical variation because they may indicate a manufacturing process problem, and the aim is to detect ``assignable causes of variation" as soon as possible, while avoiding false alarms using a statistical monitoring scheme. We consider both of what is called ``Phase 1" and "Phase 2" SPC. Figure \ref{fig:pipeline} shows a ``pipeline" of the methods we propose in this paper, starting from scanned data from the surface of an object, estimation of the Laplace-Beltrami operator of the surface and its spectrum, to SPC methods for Phase I and II, and finally, post-alarm diagnostics.

The paper is organized as follows. Section \ref{sec:2} first reviews some preliminary mathematical concepts in order to present the new differential-geometric SPC methods, including the main concept we will use, the Laplace-Beltrami (LB) operator, and its spectrum. 
The spectrum of the LB operator needs to be estimated from data, and {S}ection \ref{sec:3} discusses methods to do so. 
 Section \ref{sec:4} presents the main practical results, where a specific distribution-free multivariate chart is used to monitor a process with respect to an in-control reference data set (i.e. ``Phase II" in SPC) using the spectrum of the estimated discrete LB operator of a sequence of scanned parts. We show how the spectral methods have greater sensitivity to detect  out of control conditions in a discrete-part manufacturing process than either using the deviations from nominal ICP  method sketched above or using instead earlier SPC approaches. A post SPC alarm diagnostic is presented in {S}ection \ref{sec:5} where we investigate the use of ICP to localize the occurrence of defects on a part.  To make the presentation of our SPC proposal complete, {S}ection \ref{sec:6}  discusses a method based on the spectrum of the LB operator for the first phase (``Phase I") when monitoring a process in the absence of prior in-control data. The paper ends with conclusions and some directions for further research. The Appendices in the Supplementary Materials (on-line) contain proofs and derivations, further discussion about the relation between the LB operator and both the heat equation and the combinatorial Laplacian in networks, and a discussion about the applicability to SPC of other intrinsic, spectral distances popular in Computer Graphics and Manifold Learning that have received recent attention in the literature. Supplementary materials also include Matlab code that implements our methods and the datasets used.

\section{PRELIMINARIES}
\label{sec:2}
In this section we first define the  concepts we use in the sequence, in particular, those leading to the definition of the {\em Laplace Beltrami operator} (definition 7 below) which is our main object of study. For more on these definitions and concepts see, e.g,., \cite{Kreyszig} and \cite{ONeill} and Appendix \ref{AppD} (supplementary materials, on-line). In Differential Geometry, one begins with properties affecting the {vicinity} of a point on a surface and deduce{s} properties governing the overall structure of the surface or manifold  under consideration. 

The preliminary concepts in this section apply to either surfaces or volumes, from which we assume a scanner takes sample measurements that constitute the datasets to be used. The surfaces or volumes are instances of a $k$-dimensional manifold ($k=2$ or 3, respectively) contained in 3-dimensional Euclidean space. Informally, a manifold $\mc M$ is a $k$-dimensional space that resembles $\mathbb{R}^k$ (Euclidean space) on small domains around each of its points (we will assume $\mc M$ is compact and connected). The manifold hypothesis, useful if true in machine learning and in engineering data analysis, indicates that high ($n$-) dimensional data frequently lie on or near a lower, often curved $k$-dimensional manifold, where $k < n$. In so-called Riemannian manifolds, smooth manifolds with an inner product, we can measure distances, areas, volumes, etc. If the {\em manifold hypothesis} holds in a dataset, we say the {\em intrinsic} dimension of the data is $k$, and that the data manifold is {\em embedded} in a{n} $n$-dimensional {\em ambient} space. For manifold data, any $n$-dimensional point $\bm x$ can be completely described by defining $k$ local (intrinsic) coordinates or ``parameters" $x^1, x^2,...,x^k$. For instance, consider a {\em parametric curve} in $\mathbb{R}^3$, such as the helicoidal curve $\bm p(t) = \left(
x(t) = \cos(t),
y(t)=\sin(t),
z(t)=t
\right)
$, 
where $t \in D \subset \mathbb{R}$ is the curve coordinate or parameter. Here the intrinsic dimension of the manifold is $k=1$ (i.e., a 1-manifold) with $x^1 = t$, while the ambient space dimension is $m=3$.  In this paper, we will mostly consider data sampled from surfaces or 2-manifolds of manufactured parts, although all our methods are extendable to the case of 3-manifolds, i.e., voxel (volumetric) data. 

    {\bf Definition 1.} Any property of a manifold $\mc M$  that can be computed without recourse to the ambient space coordinates, and instead is computed using only the intrinsic or local manifold coordinates $x^1,...,x^k$ is said to be an {\em intrinsic geometrical property}, or  simply, an {\em intrinsic property}, of $\mc M$. 
    
Thus, to describe data points on a surface, such as geographical data on Earth, we need only two coordinates $x^1, x^2$, so the surface of a sphere is intrinsically 2-dimensional.

{\bf Definition 2.} Any geometrical property of an object that remains constant after application of a given transformation is said to be {\em invariant} with respect to that transformation. 

 Intrinsic geometrical properties of a manifold in Euclidean space are invariant with respect to {\em rigid transformations} (rotations and translations, but not dilations), but the opposite is not true.     The SPC methods we will present are intrinsic, and therefore invariant, and it is thanks to these properties that the groupwise part registration problem can be avoided.
An instance of Definition 2 are Euclidean distances between points in a configuration matrix $\bm X$, which are invariant to  rigid transformations but they are evidently not intrinsic. An instance of an intrinsic property is the geodesic distance between two points located on a manifold $\mc M$, which is therefore also invariant with respect to rigid transformations. In Appendix \ref{sec:7} (supplementary materials, on-line), we discuss intrinsic distances other than the geodesics and their potential use for SPC.

A classic result in Differential Geometry indicates that intrinsic geometrical properties of a manifold depend only on the so-called {\em first fundamental form} of $\mc M$, a quadratic form defined next for 2-manifolds (surfaces), the case we concentrate in this paper. Consider a parametric surface ${\bf p}(u,v) = (x(u,v), y(u,v), z(u,v))'$, $(u,v) \in D \subset \mathbb{R}^2$ (here $u=x^1, v=x^2)$ which defines $\mc M$, a Riemannian 2-manifold, i.e., a smooth (so derivatives can be computed) surface. Define the surface differential vectors at ${\bf p}(u,v)$ as:
\[ {\bf p}_u = \frac{\partial {\bf p}(u,v)}{\partial u} = \left(
\frac{\partial x(u,v)}{\partial u},
\frac{\partial y(u,v)}{\partial u},
\frac{\partial z(u,v)}{\partial u}\right)' \quad
\]
and
\[\quad {\bf p}_v = \frac{\partial {\bf p}(u,v)}{\partial v} = \left(
\frac{\partial x(u,v)}{\partial v},
\frac{\partial y(u,v)}{\partial v},
\frac{\partial z(u,v)}{\partial v}\right)'.\] 
Now define a parametric curve on $\mc{M}$, $
\alpha(t)= {\bf p}(u(t), v(t))$ such that 
$ {\bf p}(u_0,v_0)=\alpha(0)$ and use the chain rule:
\[
\frac{d \alpha(t)}{dt} = \frac{\partial {\bf p}}{\partial u} \frac{d u(t)}{dt} + \frac{\partial {\bf p}}{\partial v}\frac{d v(t)}{dt} =
 {\bf p}_u \frac{d u(t)}{dt} + {\bf p}_v \frac{d v(t)}{dt}.
\]
Finally, take the inner product (borrowed from $\mathbb{R}^3$): 
\begin{eqnarray*}
\RN{1}_p\left(\frac{d \alpha(t)}{dt} \right) &=& \la \frac{d \alpha(t)}{dt}, \frac{d \alpha(t)}{dt} \ra  \; = \; \mbox{($ds)^2$}
 = g_{11} \left( \frac{d u(t)}{dt}\right)^2 + 2 g_{12}  \frac{d u(t)}{dt}  \frac{d v(t)}{dt}+ g_{22} \left( \frac{d v(t)}{dt}\right)^2
\end{eqnarray*}
where: $
g_{11} = {\la {\bf p}_u, {\bf p}_u\ra} \quad
g_{12} = {\la {\bf p}_u, {\bf p}_v \ra} \quad
g_{22} = {\la {\bf p}_v, {\bf p}_v \ra}
$. We then have the following.

{\bf Definition 3.} The quadratic form $\RN{1}_p(\alpha(t)') = (ds)^2$  is called the {\em first fundamental form} of the parametric surface ${\bf p}(u,v)$ describing $\mc M$. It provides a means to measure arc lengths, areas and angles on $\mc M$. It defines a new inner product for vectors on tangent planes $\la , \ra_{\mc{M}}$ and therefore,  a {\em metric} on the surface, with associated matrix (tensor) $\bm g$:
\begin{center}
\begin{tabular}{cc}
$  \la {\bf w}_1, {\bf w}_2 \ra_{\mc{M}} = {\bf w}_1^T {\bm g}  {\bf w}_2 \, \, , \, \,$
&$
{\bm g} = \left(\begin{array}{cc}
g_{11} & g_{12}\\
g_{12} & g_{22}
\end{array} \right)$.\\
\end{tabular}
\end{center}
\nd In this sense, the ambient space induces a metric, the Riemannian metric, on the manifold $\mc M$. Since $|{\bf w}| = \sqrt{\la {\bf w}, {\bf w} \ra_{\mc{M}}}$, the length of a curve segment on $\mc{M}$ is:
\[
s(t) =  \int_0^t \sqrt{ds^2} \; dt .
\]
With the Riemannian metric ${\bm g}$, we can also compute differential operators acting on a function {defined} on $\mc{M}$, which are very useful for our purposes, i.e., for estimation, and therefore, statistical monitoring, of intrinsic geometrical properties of a 3D object. 

{\bf Definition 4.} The {\em gradient} of a function on $\mathbb{R}^n$ points in the direction of steepest ascent and equals to:
\[
\nabla f = \left( \frac{\partial f}{\partial x_1}, \frac{\partial f}{\partial x_2}, ....,\frac{\partial f}{\partial x_n} \right)
\]
The gradient therefore creates a vector field in $\mathbb{R}^n$.

{\bf Definition 5.} The {\em divergence} of a vector field $\bf F${$=(F_1, F_2, ..., F_n)'$} in $\mathbb{R}^n$ is given by:
\[
div \; {\bf F} = \nabla \cdot F = \frac{\partial F_{1}}{\partial x_1}+\frac{\partial F_{2}}{\partial x_2}+ ....+\frac{\partial F_{n}}{\partial x_n}
\]
\nd and measures the ``quantity" of the outward flux of $F$ from the infinitesimal neighborhood around each point $p$. This is a scalar-valued function that creates a scalar field.

{\bf Definition 6.} The {\em Laplace operator} of a twice differentiable function $f : \mathbb{R}^n \rightarrow \mathbb{R}$ is minus the divergence of its gradient field:
\[ \Delta f = - div \; \nabla f = - \sum_{i=1}^n \frac{\partial^2 f}{\partial x_i^2} \]
and measures the difference between $f(x)$  and the average $f(x)$ in a 
neighborhood around $x$. Given the second derivatives, it is a measure of curvature, and can be alternatively understood as minus the trace of the Hessian of $f(x)$. {The minus sign is for consistency with equation (\ref{LBoperator}) below.} Note how the domain of the function here is n-dimensional euclidean space. We next extend this definition to general manifolds, obtaining
the main differential-geometric operator used in the sequence, the Laplace-Beltrami operator, widely used in Computer Graphics and Machine Learning  \citep{BelkinPhD,KimmelBook,Levy,Patane,ReuterShapeDNA}.

{\bf Definition 7.} For a function $f: \mc M \rightarrow \mathbb{R}$, the {\em  Laplace-Beltrami} (LB) operator (sometimes called the {\em second differential parameter of Beltrami}, see \citealt{Kreyszig}) is defined as:
\[ \Delta_{\mc{M}} f = - div_{\mc{M}} \; \nabla_{\mc{M}} f\]
where $div_{\mc{M}} $ is the divergence taken on $\mc{M}$.  For a point defined by a parametric surface ${\bf p}(u,v)$ the following relation holds:
\be \Delta_{\mc{M}} {\bf p}(u,v) = {-}div_{\mc{M}} \; \nabla_{\mc{M}} {\bf p}(u,v) = { 2 H}  {\bf n}(u,v) \; \in \mathbb{R}^3 \label{2H}
\ee
where ${\bf n}(u,v)$ is the normal at the point ${\bf p}(u,v)$ on $\mc M$  and $H$ is the {\em mean curvature} of $\mc M$ at ${\bf p}$, which equals the average of the maximum and minimum curvatures at ${\bf p}$. This relation provides a geometric interpretation of the action of the LB operator, which can be visualized as creating a vector field of normals on $\mc M$ such that the ``height" of the normal is {twice} the mean curvature of $\mc M$ at that point.

The LB operator extends the definition of the Laplacian to functions defined on manifolds, and is an intrinsic measure of local curvature of a function at a point. Intuitively speaking, this curvature of the function needs to consider also the {\em curvature of the manifold itself}, which, contrary to the Laplacian case for a function defined on Euclidean space, is not flat. The LB operator, ``contains" the local manifold curvature. The intrinsic nature of the LB operator can be seen from defining a local coordinate system (or parametrization)  on the manifold (${\bf x}(x^1,...,x^k$), with $k=2$ for surfaces). Then, the LB operator applied to a function $f(x^1,...,x^k) \in \mathcal{C}^2$ is defined as:
\be
\Delta_{\mc{M}} f = {-}\frac{1}{\sqrt{det(g)}}\sum_{j=1}^k \frac{\partial}{\partial x^j} \left( \sqrt{det(g)} \sum_{i=1}^k g^{ij} \frac{\partial f}{\partial x^i} \right)
\label{LBoperator}
\ee
where $g^{ij}$ are the elements of $\bm g^{-1}$.  The LB operator on $f$ is therefore a function of elements in the metric tensor $\bm g$ only, and thus it is intrinsic and invariant with respect to rigid transformations. This is the key property we exploit for SPC: when considering a sequence of part surface datasets,  the spectrum (i.e., eigenvalues, see below) of the corresponding estimated LB operators can be compared directly without any need to register the parts, since it does not matter how the parts were oriented or located when measured, the LB operator remains the same. 
As discussed below, the spectrum of the LB operator contains considerable additional geometric information about the manifold, and is widely used  for this reason in both machine learning and computer graphics/computer vision. For concrete examples of the computation of the LB operator (\ref{LBoperator}) for functions $f$ defined on 3D objects, see Appendix \ref{AppD} (on-line supplementary materials).

\subsection{The Laplace-Beltrami operator spectrum and some of its properties}

We propose to work with the spectrum (collection of eigenvalues) of the estimated LB operator of a part dataset. Formally, the Laplacian eigenvalue problem is
\be
\Delta_{\mc M} f =  \lambda f
\label{Helmholtz}
\ee
sometimes called the Helmholtz partial differential equation (see \citealt[pg. 323]{Evans}, and Appendix \ref{AppD}, on-line supplementary materials), with an infinite number of pairs of eigenfunctions $f$ and eigenvalue{s} $\lambda$. The collection of eigenvalues $\{\lambda_i\}_{i=0}^{\infty}$ {(in ascending order)} is called the {\em spectrum} of the LB operator and the eigenfunctions form an orthonormal basis in $L_2(\mc M)$, see \cite{ChavelBook}.  In the particular case $\mc M$ is a circle (1-dimensional manifold) the corresponding basis functions consist of the usual Fourier harmonics $sin(k \pi x)$ and $cos(k \pi x)$.

The spectrum of the LB operator is always discrete, non-negative, and contains considerable geometrical and topological information about a manifold that can be used for shape identification. For instance, Weyl's law (see  \citealt{ChavelBook}) indicates that, for a surface $\mc M$:
\be
\lim_{i \rightarrow \infty} \frac{\lambda_i}{i} = \frac{4 \pi}{\mbox{Area}(\mc M)}
\label{eigLimit}
\ee
(thus the area of $\mc M$ can be inferred from the asymptotic slope of the spectrum; note $\lambda_i$ is proportional to the index $i$). Another result shown in a classic paper by  \cite{Kac} is
\[
\sum_{i=1}^\infty e^{-\lambda_i t} \leq \frac{\mbox{Area}(\mc M)}{2 \pi t}.
\]
Also, the spectrum contains topological information about $\mc M$. For instance, one result showing dependency of the spectrum on topological information is that for a surface without boundary \citep{YangYau},
\[
\lambda_1 \leq \frac{8 \pi(\mc G + 1)}{\mbox{Area}(\mc M)}
\]
where $\mc G$ is the genus of the surface (number of holes). 

{Scaling a surface $\mc M$ by a factor of $s$ changes the eigenvalues by $1/s^2$ (see Figure \ref{fig7}). Also, the spectrum changes in a continuous form as the shape of the manifold deforms, making it possible to monitor small shape changes through the spectrum. Furthermore, useful information can be extracted from only the lower part of the spectrum.} According to \cite{Reuter2009}, the LB operator spectrum of a manifold has more discrimination power than simpler measures like surface area. These authors provided examples of shapes with the same surface area but different spectrum. 

\section{CHARACTERIZING THE GEOMETRY OF A 3D OBJECT USING THE LB OPERATOR SPECTRUM}
\label{sec:3}

The true spectrum of very few manifolds is known. One instance where it is known is the case of a unit sphere (Figure \ref{fig7}). Repeated eigenvalues are the result of perfect symmetries in the geometry of an object and are therefore rare in practice. To characterize the geometry of general 3D objects, such as discrete parts from a manufactured process, we need to use a discrete approximation of the LB operator based on a sample of points, possibly with adjacency information, forming a mesh. Representations based on voxel data are also possible. In this paper we focus on mesh data and leave voxel laplacians for further research.


\subsection{Approximating the LB operator and its spectrum}

 The LB operator is linear, taking functions into functions.  In practice, we have no expression for the surface $\mc M$ of a part, we only have a (large) sample of points (point cloud dataset) typically returned by most 3D scanners with adjacency information (mesh or triangulation dataset). We can discretize the manifold $\mc M$ based on the data, and {functions defined on $\mc M$ are reduced to vectors whose elements are the function values on the sampled points. Then} an approximate, or {\em discrete LB operator} is obtained in the form of a matrix acting on vectors, returning a vector. One then works with the eigenvalues of the matrix which approximate the true spectrum of the {LB operator on the original} manifold. There are different ways to discretize the manifold where the data lie. In computer graphics and machine learning, it is common to work with triangulations, sometimes generated automatically by non-contact sensors. If the data are in the form of a point cloud, a graph can also {be} constructed  {by connecting} the nearest neighbors to each point. Finite-element methods (FEM) approximations of the LB operator for mesh data  \citep{ReuterShapeDNA} and for voxel data \citep{ReuterVoxel} have also been developed, which increases the practical use of this differential geometry tool.

Motivation for some of the most popular LB operator approximations used for analyzing the shape of an object comes from the theory of heat diffusion and wave propagation in Physics (see Appendix \ref{AppD}, supplementary materials, on-line). For instance, if the sensor data available have the form of a triangulation $\mc K$, the Mesh Laplacian approximation \citep{Belkin2008Discrete} is given by:
\be
L_{\mc K}^t f(p_i) = \frac{1}{4 \pi t^2} \sum_{T \in T_{\mc K}} \frac{A(T)}{3} \sum_{p_j \in V(T)} e^{-||p_i-p_j||^2/4t}(f(p_i)-f(p_j)), \quad j=1,2,..,m
\label{MeshLB}
\ee
where $T_{\mc K}$ denotes the set of all triangles in the mesh, $A(T)$ denotes the area of triangle $T$, and $V(T)$ denotes the set of vertices in triangle $T$. The parameter {$t$ comes from the heat equation, whose relationship to the LB operator is explained in Appendix \ref{AppD} (supplementary materials, on-line). Larger $t$ values imply the approximate Laplacian is considering larger areas of interest around a given point.}

We will use a recent modification of the {M}esh Laplacian (\ref{MeshLB}). We first show how the {M}esh Laplacian has a simple expression that connects it to the underlying graph Laplacian, as the next simple result, which is new as far as we know, indicates. This Lemma also applies to the {L}ocalized {M}esh Laplacian we use. The result is relevant in practice given the interpretability and applications of the graph Laplacian eigenvectors (Appendix \ref{AppD}, supplementary materials, on-line). 
  
 {\bf Lemma 1.} For a Riemannian manifold $\mc M$ sampled at {$m$ points}, the discretized approximation of its LB operator (\ref{MeshLB}) results in an $m \times m$ Laplacian matrix acting on vectors $f${$\in\mathbb{R}^m$} which can be written as:
\be
L_{\mc K}^t = D-W
\label{L}
\ee
with $W_{ij}=\frac{1}{12 \pi t^2} \sum_{T: p_j \in V(T)}A(T) e^{-||p_i-p_j||^2/4t}$ for $i=1,...,m$ and $j=1,...,m$ and diagonal matrix $D$ with entries $D_{jj}= \sum_i W_{ij}$.

For a proof, see Appendix \ref{App2} (supplementary materials, on-line). The {\em estimated spectrum} of the LB operator is then given by the eigenvalues of matrix $L_{\mc K}^t$.  \cite{Li2015} noted how the {M}esh Laplacian approximation (\ref{MeshLB}) is a {\em global} approximation, since computing it at a given point requires the integral over the whole surface $\mc M$. Instead, these authors proposed a modification of the {M}esh Laplacian that uses geodesic distances between points $p_i$ and $p_j$ (as opposed to euclidean distances) and that considers in the last sum only points within a certain radius $r$ of each point {$p_i$}, resulting in the alternative discrete Laplacian approximation:
\be\begin{aligned}
{L^t_{\mc K} f(p_i) = \frac{1}{4\pi t^2}\int_{y: d(p_i,p_j)\leq r} e^{-\frac{d(p_i,p_j)^2}{4t}} \big[f(p_i)-f(p_j)\big] dp_j}\\
{= \frac{1}{4\pi t^2} \sum_{y: d(p_i,p_j)\leq r} \frac{A(y)}{3} e^{-\frac{d(p_i,p_j)^2}{4t}} \big[f(p_i)-f(p_j)\big]}
\end{aligned}
\label{Localized}
\ee
where $A(y)$ denotes the area of the one ring neighborhood of point $y$. By writing the equation above in vector form, it is easy to obtain the Laplacian matrix in the form of equation (\ref{L}), $L^t_{\mc K}=D-W$, with $W_{ij}=\frac{1}{12 \pi t^2} A(p_j) e^{-d(p_i,p_j)^2/4t}$ for $i=1,...,m$ and $j=1,...,m$ and diagonal matrix $D$ with entries $D_{ii}= \sum_j W_{ij}$. We call this LB approximation the {\em Localized Mesh Laplacian}. It has the merit of resulting in sparser $L$ matrices, reducing storage and computational requirements, and was used in all the examples and figures shown below. 

{\bf Convergence. }Both \cite{Belkin2008Discrete,belkin2009constructing} and \cite{Li2015} show how as the triangulation $\mc K$ gets finer, their Laplacians $L_{\mc K}^t$ (\ref{MeshLB} and \ref{Localized}) converge pointwise to the continuous LB operator $\Delta_{\mc M}$ defined on a smooth manifold. \cite{dey2010convergence} further proved the convergence and stability of the spectra of the {M}esh Laplacian in \cite{Belkin2008Discrete}. The following is a Corollary of these results, important for our purposes as we use the spectrum of (\ref{Localized}), and its proof is shown in Appendix \ref{App2} (on-line supplementary materials).

{\bf Lemma 2}. For a smooth manifold $\mc M \subset \mathbb{R}^3$ with spectrum  $\{\lambda_i\}$ and for the spectrum of the Localized Mesh laplacian (\ref{Localized}), ${\{}\lambda^L_i{\}}$, we have that for fixed $i$, $|\lambda_i - \lambda^L_i| \rightarrow 0$ as $\epsilon \rightarrow 0$, $t \rightarrow 0$, and $\epsilon/t^4 \rightarrow 0$.

\begin{figure}
\centering
\includegraphics[scale=0.4]{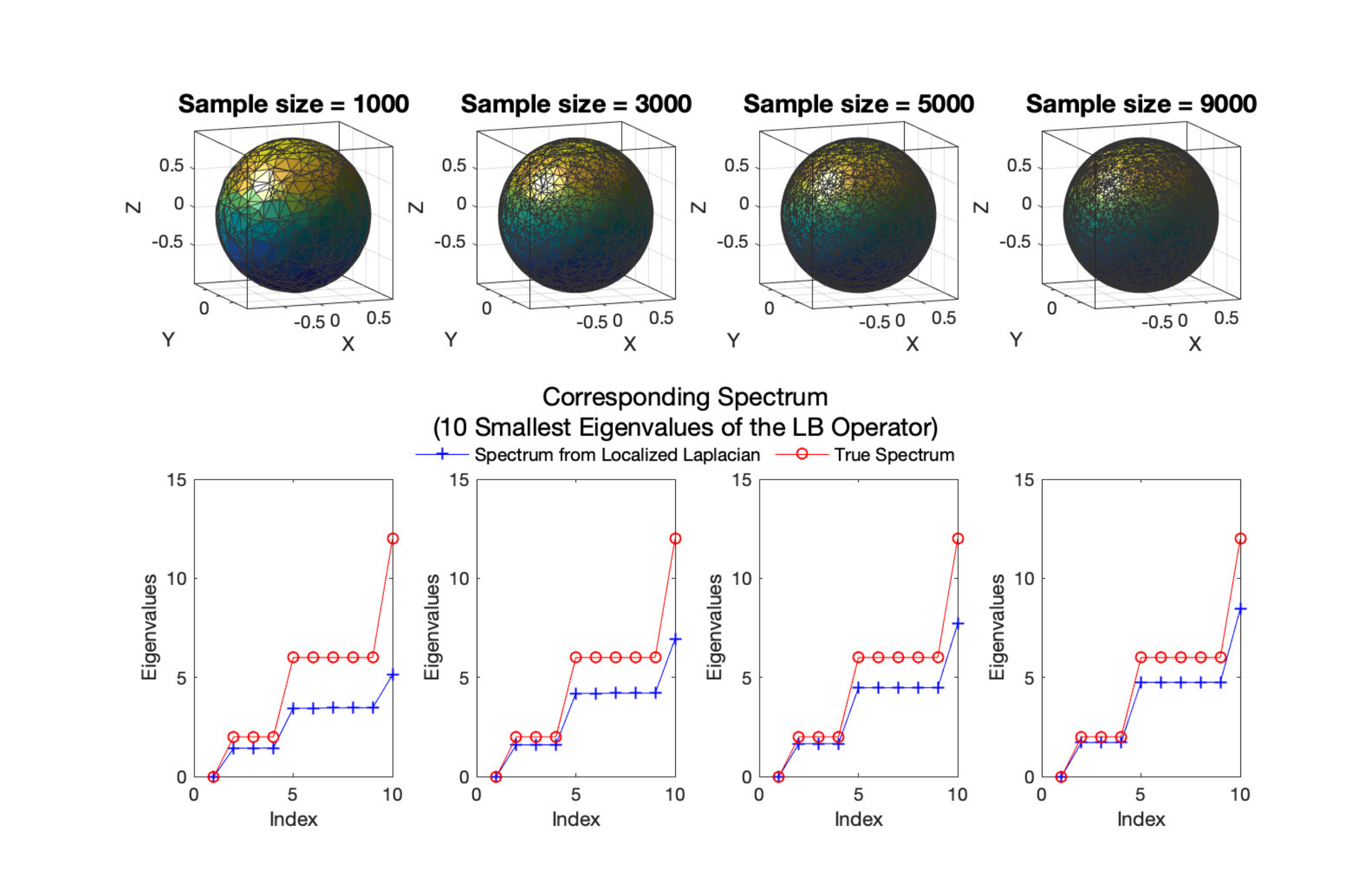}
\caption{The spectrum of the Localized Mesh Laplacian (blue) converges to the spectrum of the manifold LB operator (in red) of a sphere (one of the few objects with known LB operator spectrum) as the density of the mesh increases and approximates the manifold better, illustrating Lemma 2. Noise free meshes of different number of vertices were used.}
\label{fig9}
\end{figure}

The parameter $\epsilon$ is a function of the density of the mesh (see Appendix \ref{App2}, supplementary materials, on-line), going to zero with denser meshes. Figure \ref{fig9} illustrates how as the mesh gets denser, it approximates the underlying surface $\mc M$ better, and this implies the convergence of the spectrum of the Localized Mesh Laplacian (\ref{Localized}).

{\bf Real eigenvalues of discrete Laplacian approximations.} \cite{patane2017introduction} pointed out that many of the discretized LB operators proposed in the literature can be represented in a unified way as the multiplication of two matrices
\be
\tilde{L}=B^{-1}L
\label{symmetrizable}
\ee
where $B$ and $L$ are symmetric, positive semi-definite ($B$ is positive definite) matrices and called the mass matrix and the stiffness matrix, respectively. This means $\tilde{L}$ is {\em symmetrizable} \citep{liu2012point} and guaranteed to have real eigenvalues{, as it can be easily seen that $(\lambda, \phi)$ is an eigenvalue-eigenvector pair of (\ref{symmetrizable}) if and only if $(\lambda, B^{1/2} \phi)$ is an eigenvalue-eigenvector pair of the symmetric matrix $B^{-1/2}L B^{-1/2}$.} We note that although a symmetrizable LB approximation has therefore the desirable property of having real spectrum, not all other desirable properties of their continuous LB operator counterparts can be achieved with a discrete approximation, so there is a ``no free lunch" situation, explaining why there are so many discrete approximations proposed to the LB operator, see \cite{wardetzky2007discrete}. For instance, the eigenfunctions of the continuous LB operator form an orthogonal basis in $L_2(\mc M)$, however, the eigenvectors of the discrete LB operator approximations $\tilde L = B^{-1} L$ do not form an orthogonal basis if using the euclidean inner product, a result of the underlying meshes used for their computation not being uniform on $\mc M$ \citep{rustamov2007}. The non-uniform mesh in turn results in a non-symmetric discrete Laplacian. The eigenvectors of a discrete approximation $\tilde L = B^{-1} L$ do form an orthogonal basis but with respect to the $B$-inner product $\langle \cdot, \cdot \rangle_B$, i.e., $\langle \phi_i, \phi_j \rangle_B = \phi_i' B \phi_j = 0$ for $i \neq j$. For this reason, \cite{rustamov2007} indicates that only when the mesh is uniform one can expect a discrete Laplacian to be ``faithful" to the continuous LB operator. This indicates that mesh pre-processing techniques may be valuable, and we explore this possibility in some of the examples in Section \ref{RL} below. The next result, which is new as far as we know, assures both the Mesh and Localized Mesh Laplacians have a real spectrum (for a proof, see Appendix \ref{App2}, supplementary materials, on-line):

{\bf Proposition 1.} The Mesh Laplacian (\ref{MeshLB}) and the Localized Mesh Laplacian (\ref{Localized}) can be written as (\ref{symmetrizable}) and therefore their eigenvalues are all real.

\begin{figure}
\centering
\includegraphics[scale=0.4]{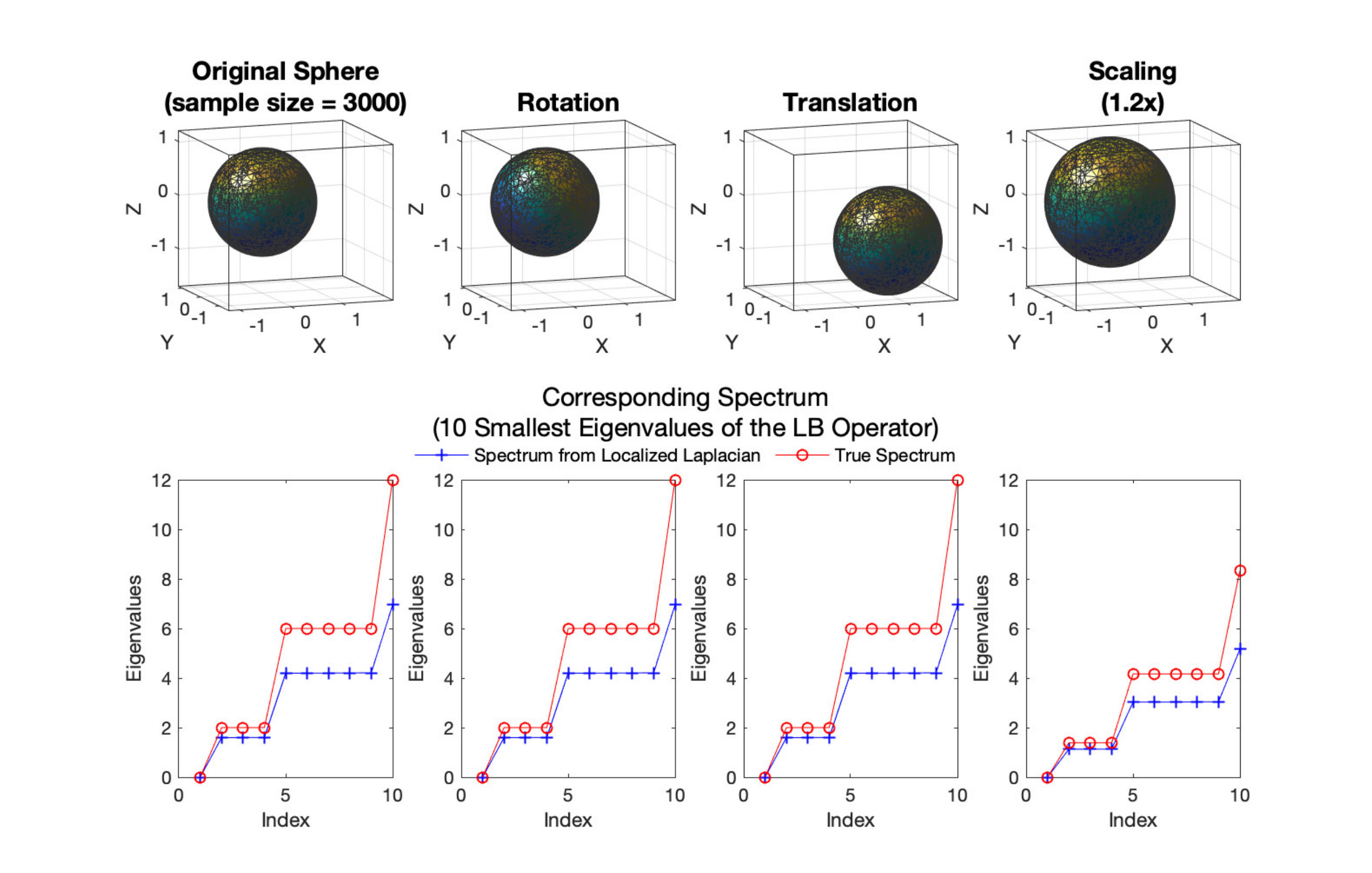}
\caption{Spectrum of the Localized Mesh Laplacian (in blue) obtained from a noise-free mesh and true spectrum of a unit sphere (in red) for different transformations. As can be seen, both spectra are invariant with respect to rigid transformations such as rotations and translations. Scaling the object by $s$ will make the eigenvalues change by $1/s^2$ (on the rightmost figures,  $s=1.2$ made the eigenvalues decrease close to 30\%).}
\label{fig7}
\end{figure}

In what follows, we use the Localized Mesh Laplacian due to its sparseness advantage over (\ref{MeshLB}).  To demonstrate numerically the behavior of this discrete Laplacian, consider Figure \ref{fig7}, which shows its first 10 eigenvalues for a mesh with 3000 points on the unit sphere. Both the true spectrum of the sphere and the spectrum of the discrete approximation are invariant with respect to rigid transformations.

\begin{figure}
\centering
\includegraphics[scale=0.4]{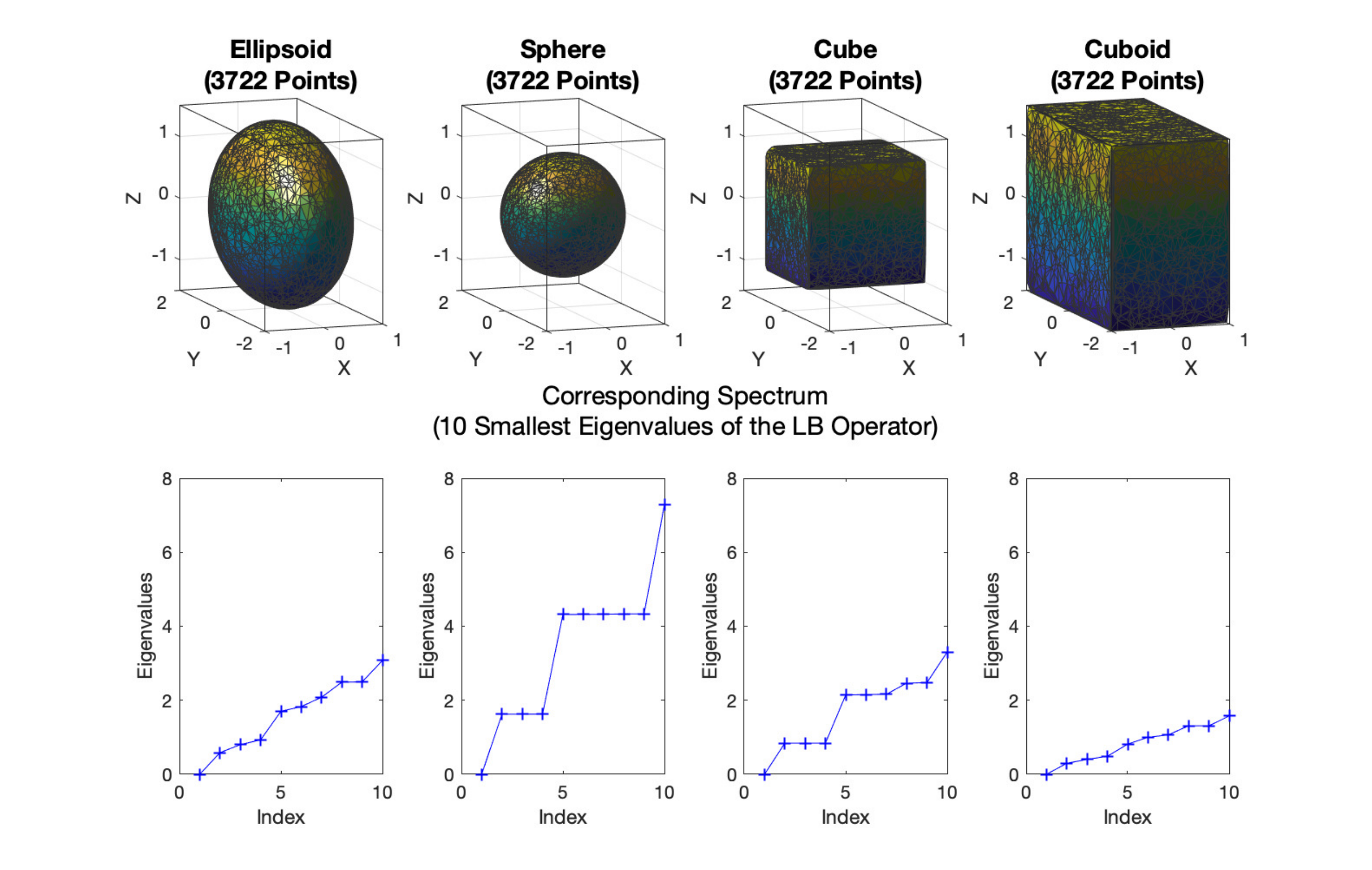}
\caption{The spectrum of the discrete LB operator contains rich geometric information about the shape of the object. Estimated Localized Mesh Laplacian spectra for different objects obtained from noise-free meshes. }
\label{fig8}
\end{figure}

\begin{figure}
\centering
\includegraphics[scale=0.4]{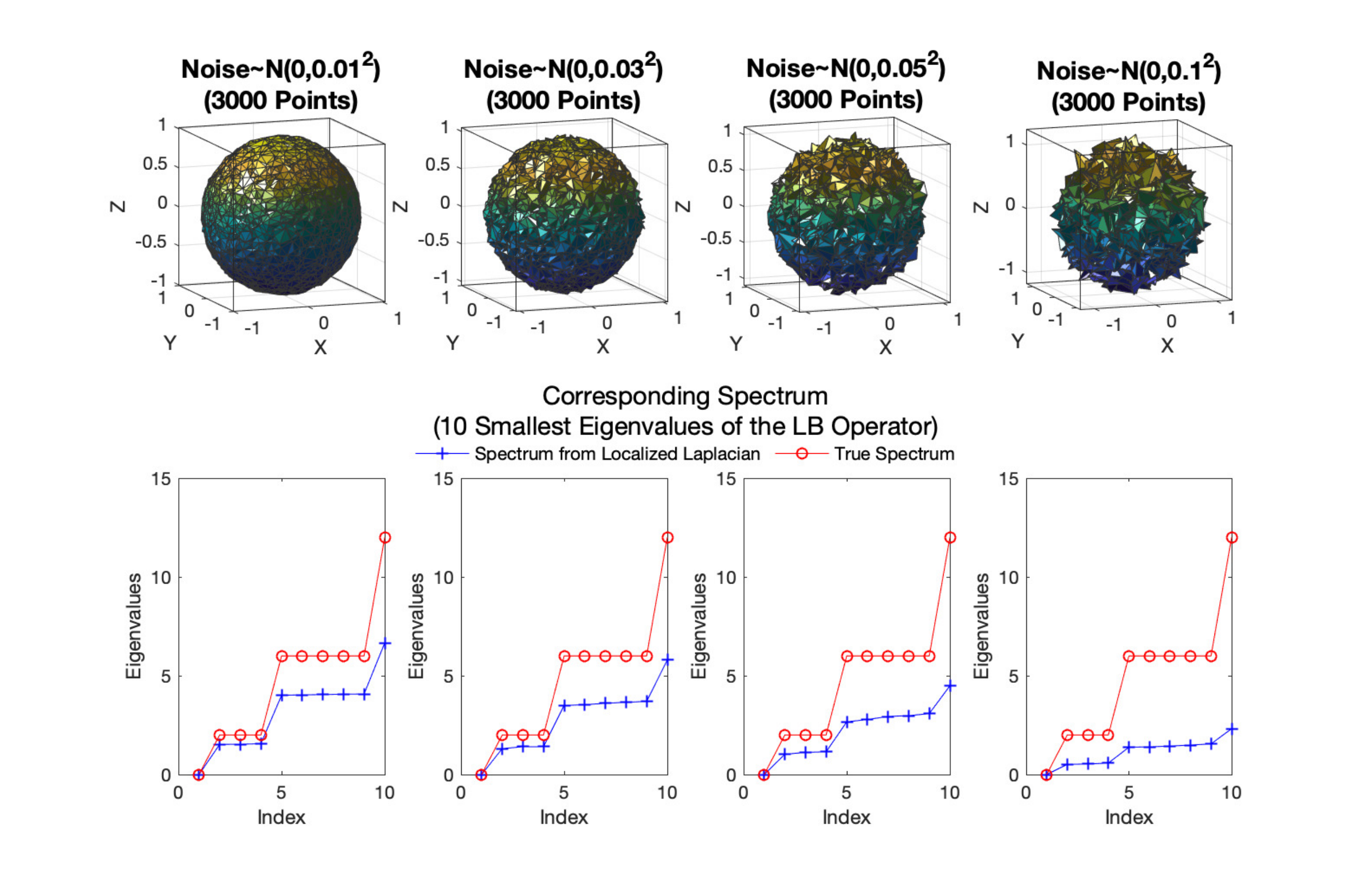}
\caption{Effect of surface noise on the Localized Mesh Laplacian (blue) compared to the true spectrum of a unit noise-free sphere (red). As the sphere loses its shape due to the severe noise, the spectrum changes, but for moderate levels of noise the lower part of the spectrum is a useful tool for SPC as proposed in this paper.}
\label{fig10}
\end{figure}

The spectra of widely different objects are considerably different, as Figure \ref{fig8} indicates.  As expected, if the noise is very high and dominates, the spectrum cannot be estimated well (Figure \ref{fig10}). For moderate levels of noise (which includes both manufacturing errors and measurement errors), we demonstrate  {in Section \ref{sec:4}} how the spectrum of the LB operator can still be used for process monitoring purposes.

\subsection{Computational complexity and stability of the discrete LB operator spectrum}

Our motivation to use intrinsic geometrical methods in SPC is to avoid the registration problem, which is a combinatorial hard, non-convex computational problem.
In our method, the computational cost of obtaining the eigenvalues of $L_{\mc K}^t$ is an $O(m)$ to $O(m^3)$ operation depending on the sparseness of the $m \times m$ matrix.

 However, extra speed can be gained from some other special properties of the LB operator besides the sparsity of its estimator. First, as it will be shown when evaluating the run-length performance of our SPC method, only the lower part of the spectrum is needed, so there is no need to compute all the eigenvalues. Second, as discussed after Lemma 2, symmetrizable LB approximations share the same spectrum with some symmetric matrices, which one can work with instead of the original approximation to take advantage of the computational benefits of symmetry, exploited by the popular Arnoldi algorithm, used in Matlab's function {\tt eigs}. The Arnoldi algorithm finds the first $k$ eigenvalue-eigenvector pairs of a sparse symmetric matrix, and has a typical complexity of $O(mk)$. Switching between the two eigenvector systems is also easy since the mass matrix $B$ is diagonal as discussed in the proof of Proposition 1. We show below how $k=15$ provides good run length performance for our LB-based SPC method, and hence, the applicability of our methods to very large datasets is possible with a desktop computer.  
 
It is relevant to point out that there are also methods to compute spectral quantities such as heat kernels and diffusion distances (discussed in Appendix \ref{sec:7}, supplementary materials, on-line) that do not require to compute the LB spectrum first \citep{Patane}.


A numerical question of importance is if the computation of the spectrum of the LB operator is stable or not. \cite{patane2017introduction} gives an analysis of the stability of this computation for each single eigenvalue $\lambda_i$, showing how the computation of eigenvalues with multiplicity one is stable numerically. 
Furthermore, our numerical experiments {(e.g. Figure \ref{fig10})} show that the lower spectrum is stable with respect to moderate levels of noise, typically encountered in manufacturing. The computation of the spectra is not stable when there are higher multiplicities \citep{patane2017introduction}, but as mentioned before, repeated eigenvalues occur due to exact symmetries which will be rare from real scanned objects, and our numerical methods did not show evidence of this kind of potential problem.

\section{USING THE ESTIMATED LB SPECTRUM AS A TOOL FOR SPC}
\label{sec:4}
Our proposal is to use the lower part of the Localized Mesh Laplacian spectrum (\ref{Localized}), i.e., the ordered spectrum cropped up to certain maximum index, obtained from scanned parts, and consider each spectrum a profile from which we derive a general statistical process monitoring technique supplemented by additional post-alarm tools to aid the localization of the defects on a part. 

\subsection{Permutation tests based on the LB spectrum}

\begin{figure}[h]
\begin{center}
{\includegraphics[scale=0.45]{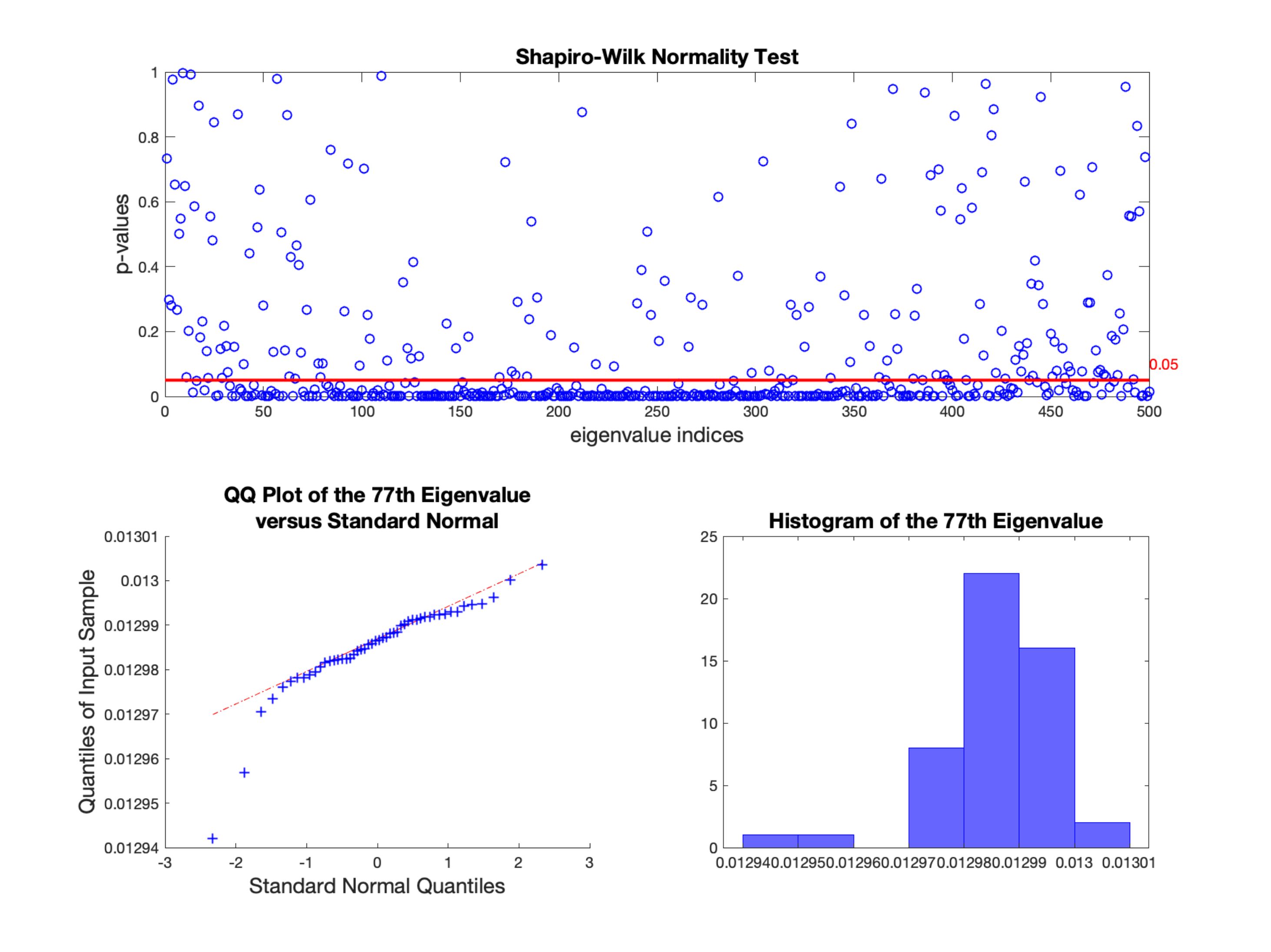}}
\end{center}
\caption{Top: p-values of the Shapiro-Wilks test of {marginal} normality for the first 500 eigenvalues from {200} realizations of the prototype part depicted in Figure \ref{fig:10} under normally distributed, isotropic noise. Bottom: QQ plot and histogram for the distribution of the simulated 77th eigenvalue. The spectra is not normally distributed even if the noise is normal.}
\label{fig12}
\end{figure}

\begin{figure}[H]
\begin{center}
\resizebox{15cm}{9cm}{\includegraphics{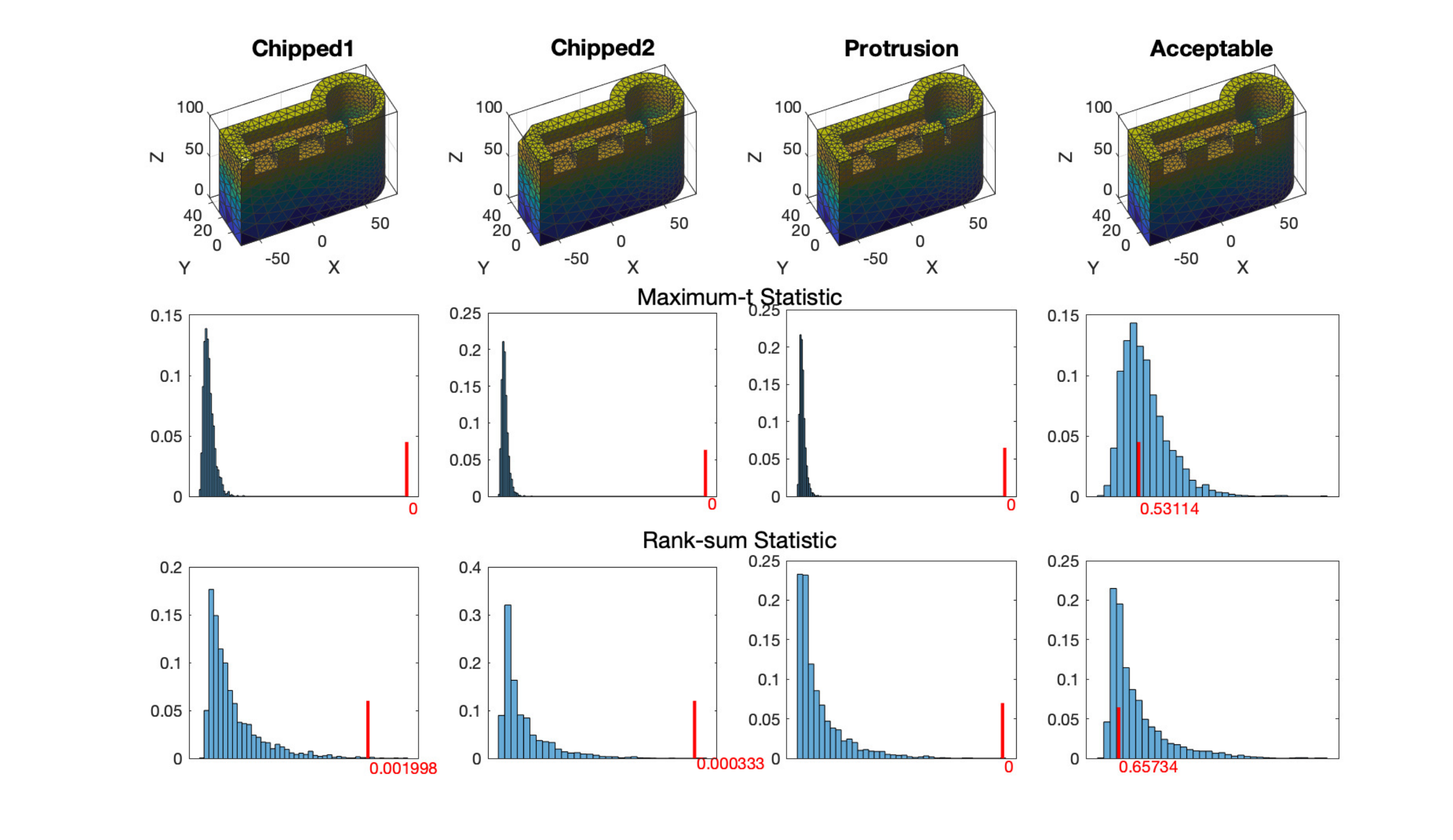}}
\caption{Permutation tests based on the statistics $t_{\max}$ (second row of plots) and the rank-sum statistic $T_0=\sum T_j^2$ (bottom row) for the difference between two groups of parts, using unnormalized eigenvalues. A sample of {5} parts of the type depicted at the top of each column was compared against a sample of size {10} of the acceptable part on the last column.  Mesh sizes ranged between 26750 and 26850, to which $N(0,{0.05^2} {\bf I}_3)$ noise was added to each vertex. The defective parts lead to strong rejections of $H_0$ (small p-values) whereas comparing acceptable vs. acceptable parts leads to failure to reject. Numbers in red (under the x-axis) are the empirical p-values based on all 3003 permutations. }
\label{fig:10}
\end{center}
\end{figure}

Empirical evidence presented by \cite{Sun2009} indicates that commercial 3D scanner noise is not Gaussian. But even for normally distributed, isotropic errors added to a surface, the estimated LB spectra are not multivariate normal. To illustrate, Figure \ref{fig12} shows the Shapiro-Wilks marginal tests of normality for each of the first 500 eigenvalues of {200} ``acceptable" parts simulated based on the prototype part  {depicted} in Figure \ref{fig:10} below to which we added isotropic N$(0, \sigma^2 \bf{I}_3)$ measurement noise. It is therefore necessary to develop distribution-free process monitoring methods for the LB spectrum.

To obtain an initial assessment of the detection capabilities of the spectrum of the LB operator, we use 2-sample permutation tests to compare the mean LB spectra between two groups of parts. Represent the spectra of the estimated LB operator (sorted from smallest to largest, up to some given {index $p$}) of each part $i$ by ${\bf X}_i \in \mathbb{R}^p$ (not to be confused with configuration matrices $\bm X$ used in section 1), and let the two samples be $\{{\bf X}_1,...,{\bf X}_m\} \sim F({\bm \mu}_0)$, when the process is in control or acceptable, and $\{{\bf X}_{m+1}, ...,{\bf X}_{m+n}\} \sim F({\bm \mu}_1)$ for parts that have an out {of} control condition. Given the non-normality of the spectrum data,  {in our preliminary tests} we use a nonparametric, distribution-free permutation test for $H_0: {\bm \mu}_0= {\bm \mu}_1$ using {two different types of statistics: the maximum t-statistic \citep{ReuterVoxel} and the} component-wise Wilcoxon rank-sum statistics \citep{Chen2016}. {The maximum t-statistic is defined as
\be
t_{\max} = \max_{1\leq j\leq p}\frac{|\sum_{i=1}^m X_i(j)/m - \sum_{i=1}^n X_{m+i}(j)/n|}{SE_j}
\label{maxT}
\ee
Here $X_i(j)$ is the $j$th element of $X_i$ (in our case,  {the} $j$th eigenvalue of part $i$), and $SE_j$ is the pooled standard error estimate of the $j$th eigenvalue
\[
SE_j = \frac{\sqrt{(m-1)\sigma_{1,m,j}^2+(n-1)\sigma_{m+1, m+n, j}^2}}{\sqrt{\frac{1}{m}+\frac{1}{n}}}
\]
where $\sigma_{i,k,j}$ is the standard deviation of the $j$th element in sample $\{{\bf X}_i,{\bf X}_{i+1},...,{\bf X}_k\}$.}

 {For specifying the} component-wise Wilcoxon rank-sum statistic, denote the rank of the jth eigenvalue from the ith part, with respect of the pool of $m+n$ parts, as $R_{ji}$ ($i=1,...,m+n$ and $j=1,...,p$) and define:
\be
{T_j = \frac{\sum_{i=1}^m R_{ji} - E[\sum_{i=1}^m R_{ji}]}{\sqrt{\mbox{Var} (\sum_{i=1}^m R_{ji})}}, \quad \quad j=1,...,p}
\label{Tj}
\ee
where $\sum_{i=1}^m E[R_{ij}] = m(m+n+1)/2$ and Var$(\sum_{i=1}^m R_{ij}) = m n (m+n+1)/12$ (see Appendix \ref{App3}, on-line supplementary materials, for a proof). The test statistic is formed by combining the $T_j${'s} using $T_0=\sum_{j=1}^p T_j^2$, with a large value of the test statistic leading to rejection of $H_0$.

Both statistics $t_{\max}$ and $T_0$ are intuitive, as there is no  {explicit mapping available} from the eigenvalues of the LB operator to the manifold $\mc M$, and differences in {\em all} the first $p$ eigenvalues should be considered jointly as evidence of a difference between the two objects.



Figure \ref{fig:10} shows the results of permutation tests between two samples of a prototypical part of realistic size, with the number of points per part varying between 26750 and 26850 points (it is important to point out that real datasets from a non-contact sensor will not have identical number of points/part). The prototype part is typical of an additive manufacturing process, and the 3 types of defects we consider below, two types of ``chipped" corner parts and a part with a ``protrusion" in one of the top edges are also typically encountered in an AM process. The first sample (with sample size 10) is  a group of acceptable parts, while the second sample (with sample size 5) is a group of parts of the type shown on top of each column, where the first three parts have different types of defects (chipped corners or a protrusion) while the last one is acceptable in the sense of being equal the CAD design of the part plus isotropic white noise. In this exercise, we added isotropic $N(0, 0.05^2 {\bf{I}_3})$ noise to all points of all parts, defective and non-defective. The second row of plots in the figure are results from the permutation test using the maximum t-statistic {as in (\ref{maxT})}, while the last row is using the Wilcoxon rank-sum statistic $T_0=\sum_j T_j^2$ with $T_j$ as in (\ref{Tj}), both using $p=500$. In each of the small figures showing the test results, the blue bars are the empirical pdf's of the test statistic from all permutations, while the red line indicates the observed value of the test statistic, $t_{\max}$ or $T_0$ . Both the maximum-t and rank-sum tests are one-sided. The red numbers under the red lines are the estimated p-values for the corresponding tests, defined as the number of permutations with more extreme (larger) values than the observed  test statistic, divided by the total number of permutations (3003 for sample sizes 10 and 5, note this is an exact permutation test).

\vspace{-0.5cm}We repeated the experiments in Figure \ref{fig:10} using eigenvalues normalized by {surface} area, given that from (\ref{eigLimit}) they relate inversely to the surface area of $\mc M$. For small, localized defects on the surface of a part, the detection capabilities of the LB spectrum using either normalized or unnormalized are very similar, and therefore the normalized spectrum results are not shown here. Since we wish to detect changes in shape and in size, and not only in shape, we suggest using the {\em unnormalized} spectrum, which will be used in the following sections.



\begin{figure}[H]
\begin{center}
\resizebox{15cm}{9cm}{\includegraphics{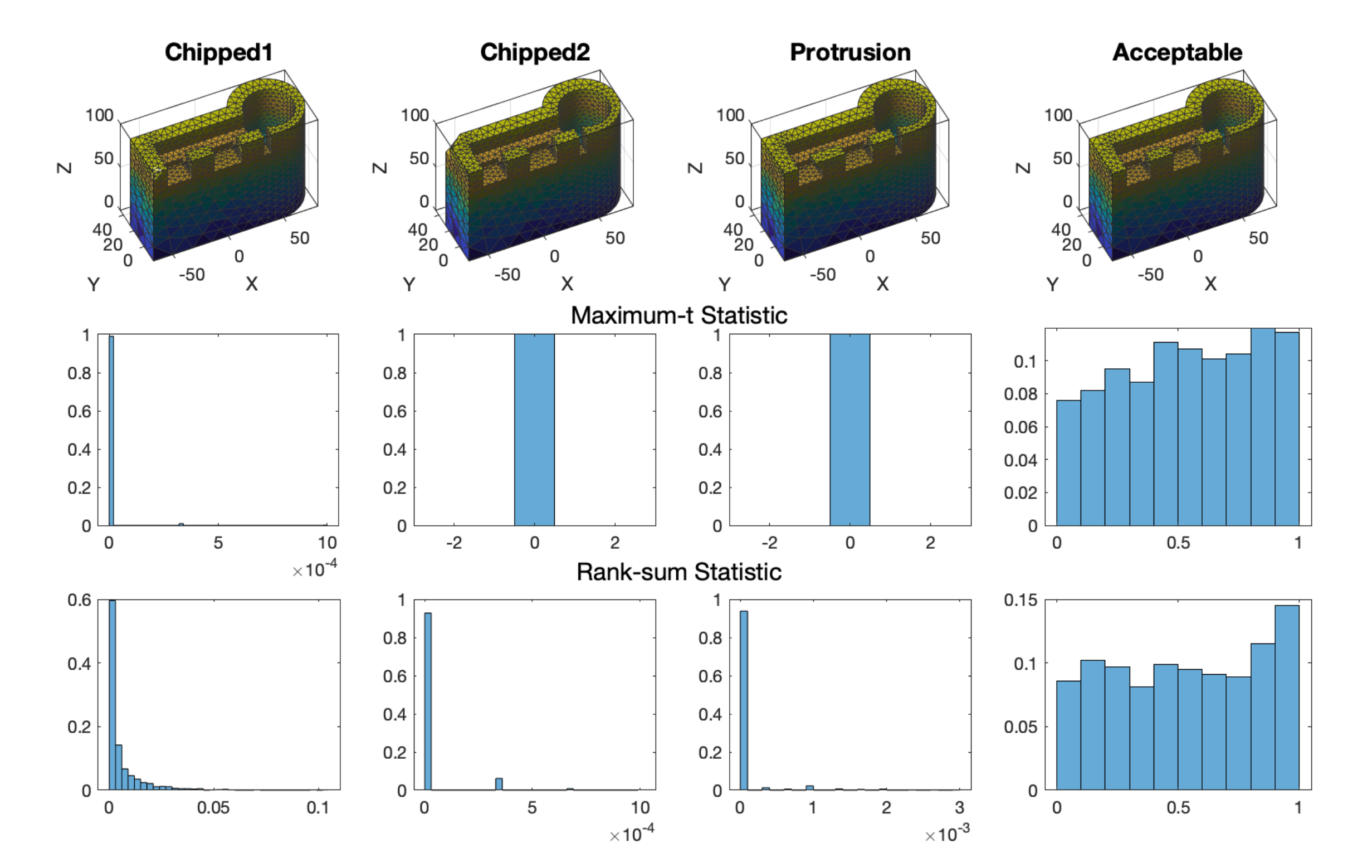}}
\caption{P-value distributions of the permutation tests of Figure  \ref{fig:10} (mesh sizes ranging from 26750 to 26850 vertices) based on the unnormalized eigenvalues.  In the first three columns, groups of defective parts of the type indicated are compared against groups of acceptable parts. Results are shown for $t_{max}$ (second row) and $T_0$ rank-sum statistics (bottom row). On the last column, acceptable parts are compared against other acceptable parts, and the p-value distribution follows a near-uniform distribution, as expected.}
\label{fig:12}
\end{center}
\end{figure}

Figure \ref{fig:12} shows the distributions of the estimated p-values (as defined above) when the permutation test procedure is repeated 1000 times. In these figures, the last column of plots shows the case when both   groups of parts consist of acceptable parts, i.e., the null hypothesis is true. As it is easy to show, in such case the theoretical p-value should follow a standard uniform distribution, and this is approximately the case  {in} the depicted histograms of p-values. In the other cases, when we are testing defective parts against acceptable parts, it is desired to have p-values very close to zero, as it is indeed the case. 

These comparisons indicate the potential for using the LB spectrum and permutation tests for shape difference detection. These are not however, an SPC scheme, since for on-line, ``Phase II" monitoring we require a sequential test, as further discussed next, and a way to initialize the scheme, or ``Phase I" SPC, which we detail in Section \ref{sec:6} . 



\subsection{On-line SPC scheme (``Phase II")}
\label{sec:4.2}

Given the non-normality of the LB spectrum, for on-line statistical control we recommend to use a multivariate permutation-based control chart. We have used, with some modifications as discussed below,  \cite{Chen2016} distribution-free multivariate exponentially-weighted moving average (``DFEWMA") chart, and applied it to the lower part of the estimated LB spectra. {Following their notation, suppose we have $m_0$ parts from Phase I and $n$ parts from Phase II, labeled as $\{-m_0+1, -m_0+2, ..., 0, 1, ..., n\}$.} Consider the set of the $j$th component of the vectors one wishes to monitor (eigenvalues of the LB spectra, in our case), taken from observation $k$ to the most recent observation $n$, $\mathcal{X}_{k,j}^n= \{X_{jk},...,X_{jn}\}$. We wish to test the equality of the location of the samples $\mathcal{X}_{-m_0+1,j}^{n-w}$ and $\mathcal{X}_{n-w+1,j}^n$, that is, the in-control observations compared to the most recent observations in a ``window" of $w$ observations.

The idea of the DFEWMA chart is to compute the ``exponentially weighted" rank statistic $(1-\lambda)^{n-i} R_{jni}$ of the last $w$ observations among all IC observations thus far{, where $R_{jni}$ is the rank of $X_{ji}$ among $\mathcal{X}_{-m_0,j}^n$}. If these ranks are {extreme (large or small)}  {this is evidence} the process has changed from its IC state. The exponential weights give more weight to the more recent observations within the last $w$ and can be useful to detect smaller process changes.  We therefore use the statistic:
\be
T_{jn}(w,\lambda)=\frac{\sum_{i=n-w+1}^n (1-\lambda)^{n-i} R_{jni}-E\left[\sum_{i=n-w+1}^n (1-\lambda)^{n-i} R_{jni}\right]}{\sqrt{Var\left(\sum_{i=n-w+1}^n (1-\lambda)^{n-i} R_{jni}\right)}}
\label{DFEWMA}
\ee
For given $m_0$ in-control observations and a false alarm probability $\alpha$ determining the geometric{ally} distributed in-control run lengths (hence the nominal IC  average run length is $1/\alpha$), the remaining chart design parameters are thus the weight $\lambda$ and the window size $w$. In Appendix \ref{App3} (supplementary materials, on-line) we derive the following moment expressions, which consider the covariances of the weighted ranks in the sum:
\be
{E\left[\sum_{i=n-w+1}^n (1-\lambda)^{n-i} R_{jni}\right] = \begin{cases}
w\frac{m_0+n+1}{2} & \lambda=0 \\
\frac{1-(1-\lambda)^w}{\lambda}\frac{m_0+n+1}{2} & \lambda\neq0 \end{cases}}
\ee
and \begin{small}
\be
\mbox{Var}\left(\sum_{i=n-w+1}^n (1-\lambda)^{n-i} R_{jni}\right)=\begin{cases}
\frac{w(m_0+n+1)(m_0+n-w)}{12} & \lambda=0 \\
\frac{1-(1-\lambda)^{2w}}{2\lambda-\lambda^2} \frac{(m_0+n+1)(m_0+n-1)}{12}\\
~~~~~-\left(\frac{1-\lambda-(1-\lambda)^w}{\lambda^2} -\frac{(1-\lambda)^2-(1-\lambda)^{2w}}{\lambda^2(2-\lambda)}\right) \frac{m_0+n+1}{6} & \lambda\neq0
\end{cases}
\ee
\end{small}

We note these expressions are corrected from those in \cite{Chen2016} (see Appendix \ref{App3},  on-line supplementary materials, for derivations). A sequential test statistic based on $T_{jn}(w,\lambda)$ statistics is the {sum of squares} $T_n(w,\lambda) = \sum_{j=1}^p T_{jn}^2(w,\lambda)$, used by \cite{Chen2016} (maximum statistics could also be used instead). Here we report results based on $T_n(w,\lambda)$ and the moment expressions {above}.

\begin{figure}[h]
\begin{center}
{\includegraphics[scale=0.50]{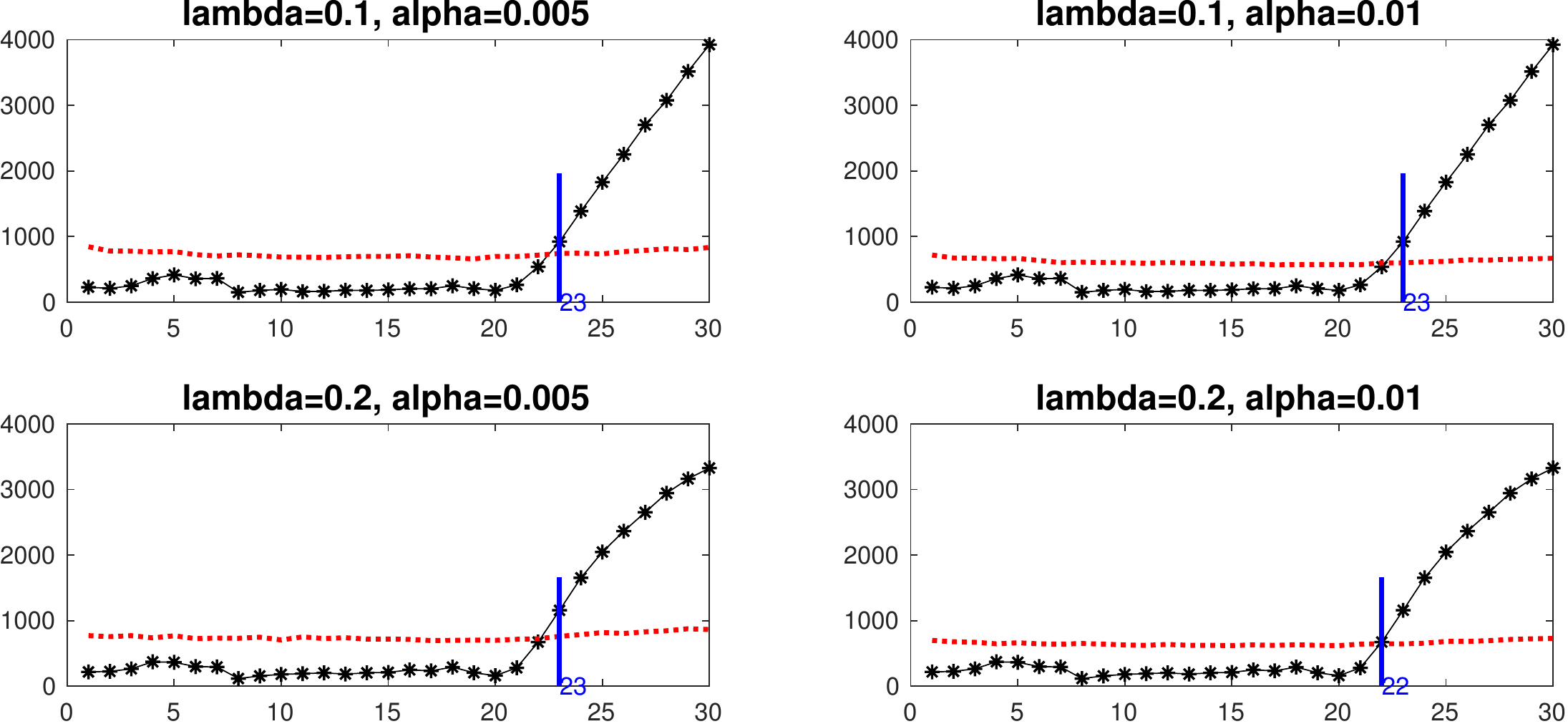}}
\end{center}
\caption{DFEWMA charts applied for ``Phase II" of a process producing the prototype part depicted in previous figures. Fifty in-control parts (with a random number of mesh {sizes} and noisy vertex coordinates) were simulated, after which the DFEWMA chart was used for on-line control. As it can be seen, the LB spectrum quickly detects the  {change in the process producing the parts}.}
\label{figSPC2}
\end{figure}

To illustrate the use of the resulting permutation chart, we simulated again parts from the CAD part model shown in Figure \ref{fig:10}. Mesh sizes varied randomly (in the range 26750 to 26850 vertices). Simulations were comprised of a ``Phase I" of 50 in-control parts (nominal plus noise), followed by an on-line ``Phase II" where defectives (parts with a {protrusion} in one of its ``teeth", see Figures \ref{fig:10}-\ref{fig:12},  {third} column) were introduced starting at part no. 21.  Figure \ref{figSPC2} shows simulations of the modified DFEWMA charts under different parameters $\lambda$ and $\alpha$. The chart has a variable control limit that depends on the observed realization of the underlying process, thanks to which it has a geometric in-control run length distribution (the main results in \cite{Chen2016} hold as well under the different test statistic (\ref{DFEWMA}) we use). In every case, detection of the out of control (OC) state occurred {within}  {the first} 3 observations.

{\bf DFEWMA chart parameter selection.-}{In common to properties of other EWMA-type of charts, smaller weight $\lambda$ leads to quicker detection of small changes, as confirmed in our simulations. Therefore, $\lambda=0.01$ was used in the following numerical results.} {As for the window size $w$,} we use $w = \min\{n, w_{\text{\scriptsize max}}\}$ with $w_{\text{\scriptsize max}}=10$ as the largest window size (as opposed to {20}, used in \citealt{Chen2016}) where $n \in \mathbb{N}$ is the number of currently available Phase II parts.

\cite{Chen2016} recommend that when the number of Phase II parts is smaller than 5, some Phase I parts should be included in the window to keep a smallest window size of 5 and prevent the number of all possible permutations from being too small, resulting in an inaccurate empirical distribution and therefore an inaccurate critical value. However, they also mentioned that {bounding} $w$ this way may reduce the detection power when a location shift occurs at the beginning of Phase II, which was found to be the case in our run-length simulations. When the smallest window size is set to 5, the DFEWMA chart needs at least 3 out-of-control parts to signal, which means the majority of the parts in the window needs to be out-of-control even if the variables are able to evidently reflect the change right away. To avoid such ``masking effect'' and truly present the detection ability of our method, while keeping the nice properties of the DFEWMA, we decided  not to include Phase I parts in the window and allow the smallest window size, $w_{\text{\scriptsize min}}$, to be 1. To keep an adequate empirical distribution, we set the number of Phase I parts, $m_0$, to 100, so that for the most extreme case, when there is only 1 part in Phase II, the number of possible permutations is ${101 \choose 1}$, providing enough permutations to form the empirical distribution. From our simulation results, this modification works well and reduces the smallest possible ARL from 3 to 2.

\vspace{-0.3cm}\subsection{Run length behavior}
\label{RL}
\begin{figure}[htbp]
\begin{center}
\includegraphics[scale=0.6]{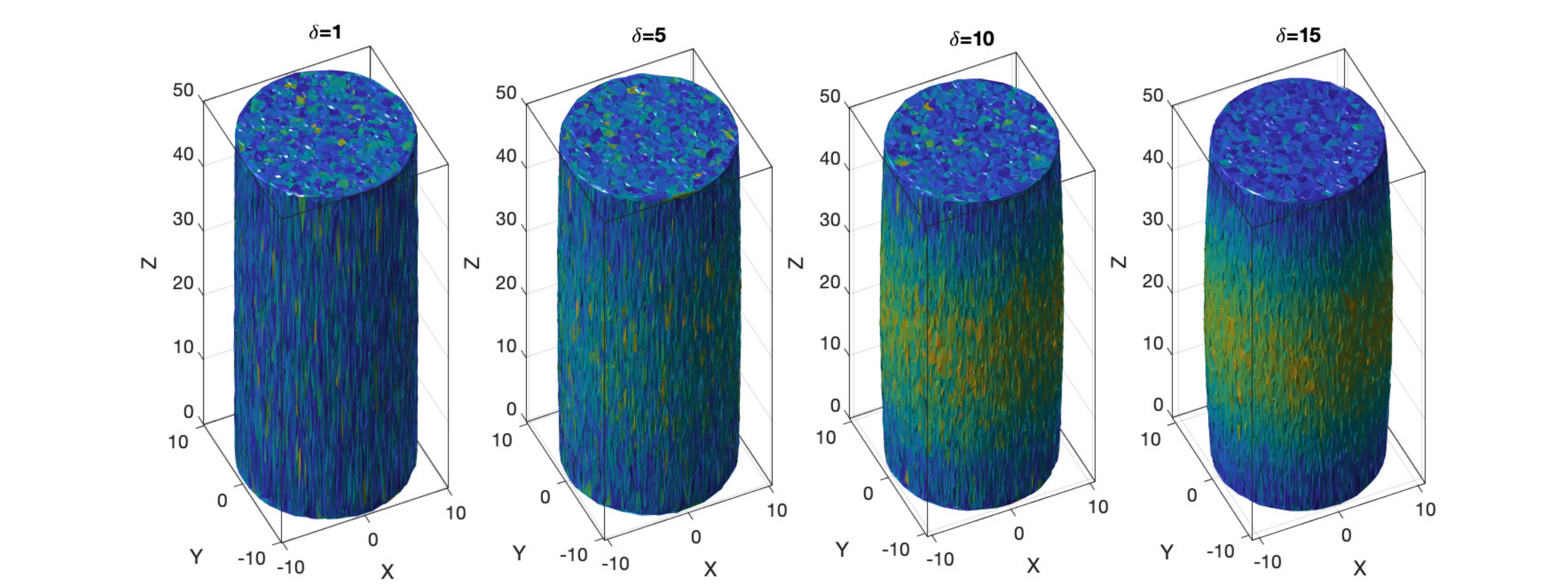}
\caption{Cylindrical parts used in the run length study for different values of the non-cylindricity parameter $\delta$ (``barrel" type defect for larger $\delta$).  Lighter areas indicate greater deviations from the nominal (CAD) model, obtained with the ICP post-alarm diagnostic described in Section \ref{sec:5}.}
\label{fig:diagnostic}
\end{center}
\end{figure}

To gain a more complete sense of the effectivity of the SPC chart, we conducted  {a run length analysis based on simulation of cylindrical parts of increasingly more deformed shape, with parts acquiring a more ``barrel-like" shape as an OC parameter $\delta>0$ is {increased}, to permit computation of out of control run lengths parametrized in a simple way (see Figure \ref{fig:diagnostic}). This is one of the typical out of control signals \cite{colosimo2014profile} discussed in the fabrication of cylindrical parts in a lathe process. We also conducted a run length analysis simulating realizations of the prototype part (and its defect types) shown in Figure \ref{fig:10}. Given that a run length analysis implies computation of thousands of LB spectra, to avoid long simulation times, we used smaller mesh sizes, with $1995\sim 2005$ points for the cylindrical parts and $1675\sim 1680$ points for the parts in Figure \ref{fig:10}. We also permuted the already simulated parts instead of simulating new parts for new replications to further reduce the computational cost of the simulation while keeping the variability of the run lengths in our analysis.} 

\begin{table}[H]
\begin{center}
\begin{tabular}{c|c|c|c|c}
\hline
\multirow{2}{*}{} & \multicolumn{2}{c|}{LB spectrum} & \multicolumn{2}{c}{ ICP objective} \\
\cline{2-5}
	& ARL & SDRL & ARL & SDRL  \\
\hline
Geometric Distribution & 20.0000 & 19.4936 & 20.0000 & 19.4936 \\
\hline	
In-control cylinder & 20.4628 & 20.2390 & 20.2492 & 19.8825 \\
\hline
In-control part & 20.1654 & 19.8863 & 20.1335 & 19.2041 \\
\hline
\end{tabular}
\caption{In-control run length performance of the DFEWMA charts applied to the LB spectrum and ICP objective. Results are obtained from 10,000 replications. Chart parameters were set at $m_0=100, w_{\text{\tiny min}}=1, w_{\text{\tiny max}}=10, \lambda=0.01,$ and $ \alpha=0.05$ which corresponds to a geometric in-control ARL of 20. The cylinder and the part in Figure \ref{fig:10} are both equal to their CAD model plus isotropic noise N$(0, 0.05^2 {\bf I}_3)$. First 15 LB operator eigenvalues were used.}
\label{ARL:IC}
\end{center}
\end{table}

\vspace{-0.5cm}{We applied the DFEWMA charts --with the corrected moments as described in Section \ref{sec:4.2}-- to the top $p=15$ eigenvalues of the LB spectrum and also  {applied it to the optimal function value returned by the ICP algorithm (\ref{ICP})}, which is a measure of the difference between two configuration matrices after discounting similarity transformations. Table \ref{ARL:IC} shows the average run lengths (ARL) and the standard deviation of the run lengths (SDRL) in the in-control processes, where we use both a perfect cylinder with radius 10 and height 50 and the acceptable part in the last column of Figure \ref{fig:10}, both with isotropic N$(0, 0.05^2 \bf{I}_3)$ noise added. As the table shows, all run lengths have nearly geometrical behaviors, which is consistent with the theorem given in \cite{Chen2016}.} 

For the  {out-of-control analysis} of the cylindrical parts, we added a first harmonic with amplitude $\delta$ times the standard deviation of the noise to the radius, so the deformed radius at height $h$ becomes $10+0.05\delta\sin(h\pi/50)$,  {adding also isotropic $N(0, 0.05^2 {\bf I}_3)$ noise to} the points. Table \ref{ARL:OC-cylinder} compares the ARL and SDRL of both methods (LB spectrum and ICP)  {as a function of the OC parameter $\delta$.} The LB spectrum is very sensitive to changes in the shape of an object, {taking only two parts to signal} the deformation of the cylinder for $\delta\geq 0.005$. On the other hand, the  ICP objective is less efficient than the LB spectrum, especially when $\delta$ is very small (note also the large SDRL values). This is because the small increase in the ICP statistic caused by a slight {local} deformation can be masked by the {overall} natural variability of the in-control process, plus the inherent variability in the ICP algorithm itself (recall that (\ref{ICP}) is a hard non-convex combinatorial problem), making it difficult for the chart to distinguish the change to an OC condition until more parts are available. However, this would not be a problem for the lower part of the LB spectrum because it is reflecting the overall shape of the parts. {\cite{Chen2016}} recommend their chart when the ratio $m_0/p$ is small, although in this case we observed good run length performance when $m_0/p=100/15$. 

\begin{table}[H]
\begin{center}
\begin{tabular}{c|c|c|c|c}
\hline
\multirow{2}{*}{} & \multicolumn{2}{c|}{LB spectrum} & \multicolumn{2}{c}{ ICP objective} \\
\cline{2-5}
	& ARL & SDRL & ARL & SDRL  \\
\hline
$\delta=0.0005$ & 10.7949 & 9.9443 & 83.2122 & 122.6474 \\
\hline	
$\delta=0.005$ & 2.0336 & 0.1851 & 39.7576 & 65.8904 \\
\hline
$\delta=0.5$ & 2.0000 & 0.0000 & 31.4895 & 51.8724 \\
\hline
$\delta=1$ & 2.0000 & 0.0000 & 5.1867 & 3.0066 \\
\hline
$\delta=2$ & 2.0000 & 0.0000 & 2.0262 & 0.1695 \\
\hline
$\delta=3$ & 2.0000 & 0.0000 & 2.0000 & 0.0000 \\
\hline
$\delta=10$ & 2.0000 & 0.0000 & 2.0000 & 0.0000 \\
\hline
\end{tabular}
\caption{Phase II out-of-control run length performance of the DFEWMA charts applied to the LB spectrum and ICP objective for barrel-shaped cylindrical parts, 10,000 replications, each with 100 IC cylinders followed by a sequence of defective  cylinders until detection.  Chart parameters  are: $m_0=100, w_{\text{\scriptsize min}}=1, w_{\text{\scriptsize max}}=10, \lambda=0.01$ and $\alpha=0.005$, corresponding to an in-control ARL of 200. First 15 LB operator eigenvalues were used, and mesh sizes varied between 1995 and 2005 points.}
\label{ARL:OC-cylinder}
\end{center}
\end{table}

We also conducted the OC run length analysis for the prototype part and defects shown in Figure \ref{fig:10}, where we consider three types of defects corresponding to the first three columns in the figure.  The ARLs and SDRLs for both methods (LB spectrum and ICP) for the three defective parts are summarized in Table \ref{ARL:OC-testpart}. In this case, the ICP method works consistently better than the LB spectrum to detect all three types of defective parts. This is because the three defects are very localized and evident to the eye, making the increase in the ICP objective function quite significant. As these three defect types are local {\em and the mesh size used is very small}, they do not change the overall shape of the part enough for the LB spectrum to quickly detect the changes in the process, particularly with the chipped part \#1, which is the smallest and more localized type of defect, with a faster detection for the protrusion defect part, which is the largest of the 3 defects relative to the mesh. To demonstrate that with larger meshes the LB spectrum would detect these types of defects quicker, we applied a pre-processing step to the meshes, utilizing the Loop subdivision method \citep{Loop}, resulting in meshes of around 5000 points instead, and this notably improves the run length performance, event though no new information is added by the pre-processing apart from the original meshes. 

For still larger meshes, the performance of the LB-spectrum will further improve. Consider Figure \ref{fig:10} where the average mesh size was 26,800 points, and the p-values of the  Wilcoxon rank-sum tests (bottom row) are significant for all three cases. This indicates that the DFEWMA chart statistic will signal faster with larger mesh sizes. We also note that chipped defect \#1 has a slightly larger p-value in Figure \ref{fig:10}, and in Figure \ref{fig:12} the distribution of its p-values has a thicker {and longer} right tale than for the chipped \#2 and protrusion defects. This shows the chipped \#1 defect is harder to detect\st{ed} by nature and explains why it results in the longest run lengths.


\begin{table}[H]
\begin{center}
\begin{tabular}{c|c|c|c|c|c|c}
\hline
\multirow{3}{*}{Defect type} & \multicolumn{2}{c|}{LB spectrum} & \multicolumn{2}{c|}{LB spectrum} & \multicolumn{2}{c}{\multirow{2}{*}{ICP objective}}\\
& \multicolumn{2}{c|}{(original mesh)} & \multicolumn{2}{c|}{(pre-processed mesh)} & \multicolumn{2}{c}{}\\
\cline{2-7}
	& ARL & SDRL & ARL & SDRL & ARL & SDRL \\
\hline
Chipped  {\#1} & 158.1168 & 182.1803 & 5.0943 & 2.7707 & 2.0000 & 0.0000 \\
\hline	
Chipped  {\# 2} & 91.3750 & 135.3055  & 4.4362 & 2.1155 & 2.0000 & 0.0000\\
\hline
Protrusion &  3.6450 & 1.7183  & 2.4341 & 0.5115 & 2.0000 & 0.0000\\
\hline
\end{tabular}
\caption{Phase II out-of-control run length performance of the DFEWMA chart applied to the LB spectrum and the ICP objective for the prototype part, isotropic uncorrelated noise. Results from 10,000 replications, each with 100 IC parts followed by a sequence of defective parts until detection. DFEWMA chart parameters were: $m_0=100, w_{\text{\scriptsize min}}=1, w_{\text{\scriptsize max}}=10, \lambda=0.01, \alpha=0.005$ (in-control ARL = 200). The first 15 eigenvalues of the LB operator were used. Mesh sizes were necessarily small (1675 to 1680 points) with pre-processing based on the Loop method resulting in close to 5000 points. }
\label{ARL:OC-testpart}
\end{center}
\end{table}

{\bf Effect of spatially correlated, non-isotropic noise.-} The run-length analysis for the barrel-shape cylinders was repeated under spatially-correlated, non-isotropic noise, as manufacturing noise may be spatially correlated on the surface of the objects depending on how the cutting tool operates on the surface. 
For this case, the defects are the same as before, so the deformed radius at height $h$ is still $10+0.05\delta sin(h\pi/50)$, where $10$, $50$, $0.05$ are the nominal radius, nominal height, and the standard deviation of noise, respectively. At each point $\boldsymbol{p}_i=\begin{pmatrix} p_{i,x}, p_{i,y}, p_{i,z}\end{pmatrix}$, non-isotropic and spatially correlated noise $\boldsymbol{e}_i=\begin{pmatrix} e_{i,x}, e_{i,y}, e_{i,z}\end{pmatrix}$ is added to the point coordinate. The covariance functions between different noise terms are:
$$Cov(e_{i,k}, e_{j,l})=\begin{cases} \sigma_1^2e^{-|p_{i,k}-p_{j,l}|/r_k} & \text{ if } i\neq j, k=l \\
\sigma_1^2+\sigma_2^2 & \text{ if } i=j, k=l \\
0 & \text{ if } k\neq l
\end{cases}$$
Here $i, j$ are point indices and $k, l\in\{x,y,z\}$ indicate the axes. To keep the same level of noise, $\sigma_1^2+\sigma_2^2=0.05^2$. {Same as in the previous example, the mesh sizes for the cylindrical parts randomly vary between 1995 and 2005 points and the first 15 eigenvalues are used.}
{Table} \ref{ARL:OC-cylinder-nonisotropic} shows the results. As it can be seen, the effect of the spatial correlation on the run length properties is negligible in the LB-spectrum method for the same levels of noise, but badly affects the chart based on the ICP objective.

{In summary, the ICP objective method is an effective method to detect evident local defects in small meshes with non-correlated noise, but as the size of the meshes grows or non-isotropic spatially correlated noise increases the registration it requires deems the method either infeasible or ineffective. On the other hand, the LB spectrum remains very sensitive to global shape changes regardless of the covariance structure of noise. For small local changes, the LB spectrum needs denser meshes to signal quickly, a condition automatically satisfied when data are obtained from non-contact sensors.}

\begin{table}[H]
\begin{center}
\begin{tabular}{ccc|c|c|c|c}
\hline
\multirow{2}{*}{$\sigma_1^2$} & \multirow{2}{*}{$r_x, r_y$} & \multirow{2}{*}{$\delta$} & \multicolumn{2}{c|}{LB spectrum} & \multicolumn{2}{c}{ ICP objective} \\
\cline{4-7}
& &	& ARL & SDRL & ARL & SDRL  \\
\hline
\multirow{3}{*}{$0.02^2$} & \multirow{3}{*}{$2.6$}  & $0.0005$ & 11.4954  &  13.8908& 198.0817 & 196.8560 \\
\cline{4-7}	
& & $0.005$ & 2.0315 & 0.1852 & 204.6340 & 200.7492 \\
 \cline{4-7}	
& & $1$ & 2.0000 & 0.0000 & 14.5372 & 17.6595 \\
\hline
\multirow{3}{*}{$0.05^2$} & \multirow{3}{*}{$2.6$}  & $0.0005$ & 8.4815 & 6.8473 & 202.2944 & 201.3962 \\
\cline{4-7}	
& & $0.005$ &2.0087  & 0.0950 & 203.9053 & 199.6954 \\
 \cline{4-7}	
& & $1$ & 2.0000 & 0.0000 & 26.6083 & 38.8395 \\
\hline
\multirow{3}{*}{$0.05^2$} & \multirow{3}{*}{$5.2$}  & $0.0005$ & 7.9981 & 6.0534 & 198.0303 & 202.7248 \\
\cline{4-7}	
& & $0.005$ & 2.0077 & 0.0886 & 194.4391 & 195.0273 \\
 \cline{4-7}	
& & $1$ & 2.0000 & 0.0000 & 28.6818 & 46.9065  \\
\hline
\end{tabular}
\caption{Phase II out-of-control run length performance of the DFEWMA charts applied to the LB spectrum and ICP objective with barrel-shaped cylindrical parts under spatially correlated, non-isotropic noise. 10,000 replications, each consisting of 100 IC cylinders followed by defective cylinders until detection. DFEWMA charts were applied to both LB spectrum method and ICP with parameters: $m_0=100, w_{\text{\scriptsize min}}=1, w_{\text{\scriptsize max}}=10, \lambda=0.01$ and $\alpha=0.005$, corresponding  to an in-control ARL of 200. First 15 LB operator eigenvalues used, mesh size varied between 1995 and 2005 points. {$r_z=16.7$ for all three cases.}}
\label{ARL:OC-cylinder-nonisotropic}
\end{center}
\end{table}

\vspace{-1cm}{\bf Comparisons versus other SPC methods for 3D data.} We finally compare the Phase II behavior of our LB spectrum method with an existing SPC method for 3D geometrical data due to  \cite{colosimo2014profile} which is based on Gaussian Processes. It should be pointed out that this is a method aimed at {\em contact sensed} data and hence assumes small, equally sized meshes with corresponding points from part to part distributed in a lattice pattern, and is a method that performs GPA registration of the points first. Their  method cannot handle the harder problem of non-contact data, where the numbers of points per part varies and points do not correspond from part to part, and would have trouble if points did not form a lattice.  Still, Table (\ref{tab:9}) shows how that our method is very competitive in these unfavorable circumstances, and, even in some cases it actually provides better run length performance.
\vspace{-0.2cm}\begin{table}[H]
\begin{center}
\begin{tabular}{c|c|c|c|c}
\hline
\multicolumn{2}{c|}{} & LB spectrum & GP$_\text{sub\_unif}$ &GP$_\text{sub\_lh}$ \\
\hline
\multirow{2}{*}{In Control} & ARL &99.8490 & 99.6890  & 100.7650  \\
 &(SDRL) &(94.3316) &  (97.4138)& (100.9427)  \\
\hline
Quadrilobe & ARL &6.2860 &4.6980  & 1.3890  \\
 $\delta=0.00185$ &(SDRL) &(2.6505) &  (3.9684)& (0.7709)  \\
\hline
Half frequency & ARL & 3.2710 & 14.1130  & 4.5090 \\
 $\delta=0.00075$ &(SDRL) &(1.4372) &  (13.4398)& (4.0450)  \\
\hline
Tapering& ARL & 61.5650 & 80.8110 & 83.1830\\
 $\delta=0.1$ &(SDRL) &(76.9659) & ( 77.6667)& (83.5048)  \\
\hline
\end{tabular}
\caption{Out-of-control run length performance comparisons of LB spectrum versus a Gaussian process (GP) model method in \cite{colosimo2014profile}. Results obtained from 1,000 replications. DFEWMA chart parameters are: $m_0=100, w_{\text{\scriptsize min}}=1, w_{\text{\scriptsize max}}=10, \lambda=0.01$ and $\alpha=0.01$, corresponding to an in-control ARL of 100. First 15 LB operator eigenvalues were used, and the mesh size is fixed to 1054 points. {Mesh pre-processing based on the Loop method is applied for the LB spectrum, resulting in around 2000 points.} Results of the GP methods were originally reported in Table 3, in \cite{colosimo2014profile}. }
\label{tab:9}
\end{center}
\end{table}

\section{POST ALARM DIAGNOSTICS}
\label{sec:5}
We have proposed a multivariate permutation SPC chart on the lower spectrum of the LB operator as a tool to detect {\em general} out of control (OC) states in the process that are not precisely defined, similarly to the role standard Shewhart charts have, as described by \cite{BoxRamirez}. Once an alarm is triggered, an investigation of the {\em specific} assignable cause that is normally carried out should include the {\em localization} of the defect on the part surface or manifold $\mc M$, a task we now describe.

{Suppose we have a part that has} triggered an alarm in the SPC charts described above and we have also have available a CAD model for the  {part being produced}. In order to localize the defect on each part, we apply the ICP algorithm to register the CAD model and  {the part that triggered the alarm (we use the ICP algorithm implemented by \citealt{Bergstrom2014})}. Upon completion, the ICP algorithm provides for each point on  {the} defective part the index of the closest point on the {noise-free CAD model}, so that {\em deviations from target} can be computed as the Euclidean distance between them (this is the minimum distance {$\min_{j=1,...,m_2}C(\bm \Gamma \bm x_{q,i}+ \bm \gamma, \bm x_{p,j})=\min_{j=1,...,m_2} ||\bm \Gamma \bm x_{q,i}+ \bm \gamma - \bm x_{p,j}||$} in problem (\ref{ICP}), with ${\bm X}_q$ and ${\bm X}_p$ being the {OC part} and the {CAD model}  {respectively}).

\begin{figure}[htbp]
\begin{center}
\includegraphics[scale=0.46]{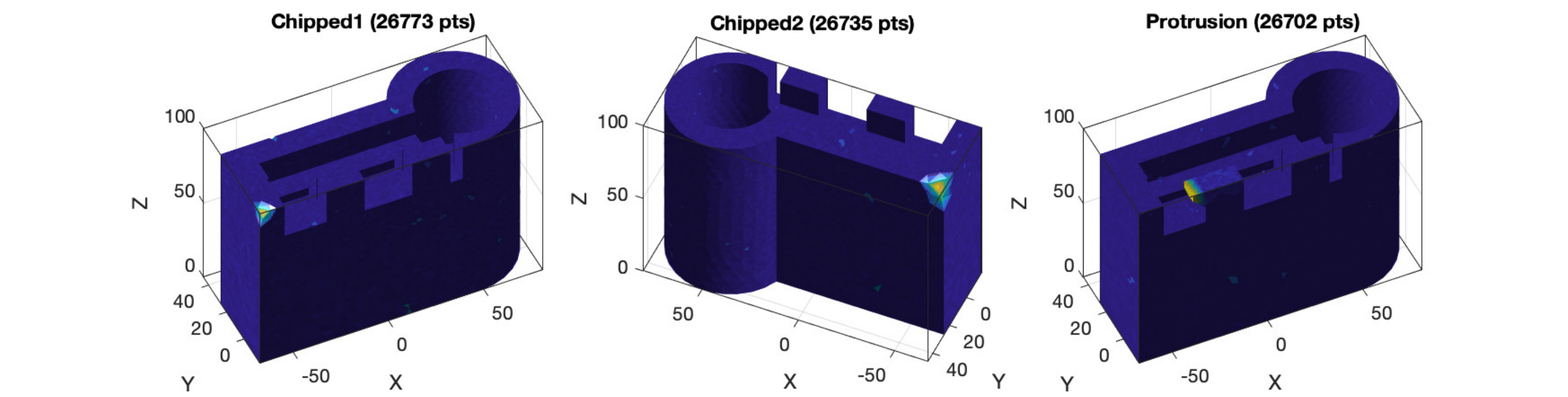}
\caption{ICP post-alarm diagnostic to localize the occurrence of a defect on the surface of a part. Lighter areas indicate greater deviations from the nominal (CAD) model.}
\label{fig:15}
\end{center}
\end{figure}

{Figure \ref{fig:15} shows three different locally defective parts, each with different number of points and with isotropic errors $N\sim (0, 0.05^2)$ added to all three coordinates.} {We} color each point on the {OC part} proportionally to these deviations, with lighter colors corresponding to larger deviations. As it can be seen, the location of each of the 3 defects on a part, the two parts with chipped corners and the part with a protrusion in one of its ``teeth", is very accurately identified. We suggest to conduct this ICP localization diagnostic after each SPC alarm.

 {When {the} change in shape {is rather small in the sense that point-wise deviations do not increase sharply}, the ICP localization diagnostic will not work as well as in the case  {where} the change of shape is very evident. To illustrate,} Figure \ref{fig:diagnostic} shows {four} out-of-control cylindrical parts with increasing $\delta$ values. The number of points varies from part to part and the same isotropic noise we have been using is added. Each part is color coded by the deviation from  {CAD} target and lighter colors means larger deviations. The global deformation of the cylinders  {is strongest along the ``waist'' of the cylinder, and is only detectable by the ICP registration when  $\delta$ is large enough to be quite evident to the eye}.  This is consistent with our findings in the OC run length analysis, where the  ICP objective is more effective with relatively larger $\delta$ values. A similar ``defect localization" could be performed with a registration method based on other distance functions between points, such as spectral distances, see Appendix \ref{sec:7} (supplementary materials, on-line).\\
 
An additional diagnostic worth mentioning is the nonparametric estimator of the change point suggested by \cite{Chen2016}, which works from the test statistics used by their DFEWMA chart {and} evidently can be used in the present situation as well, if desired, to determine the first part in the sequence that needs to be investigated with the ICP diagnostic.


\section{A PERMUTATION-BASED SPC SCHEME FOR ``PHASE I"}
\label{sec:6}

{Although the chart by \cite{Chen2016} is self-starting, these authors recommend to perform a ``Phase I" of $m_0$ parts to avoid masking effects in their chart if a problem occurs very early after startup, with recommended values of $m_0$ of at least 50. If parts are expensive, it is important to have an additional scheme that can detect an out of control process within this period with high probability. We therefore suggest using the distribution-free multivariate chart proposed by \cite{capizzi2017phase},  {available in their} R package {\bf dfphase1},  {applied to the lower spectrum of the LB operator}. Similar to the DFEWMA chart,  {these authors} used rank related statistics, so their method is more sensitive to small rather than large changes in the process.  {
{Following \cite{capizzi2017phase}, the performance of the chart is evaluated based} on the false alarm probability (FAP) for an in-control process and the alarm probabilities for out-of-control scenarios.} Given the ``Phase I'' dataset consisting of the cropped spectra, we first discard the first  eigenvalue of each part because  {it is theoretically zero (non-zero first eigenvalues from the estimated LB operator are pure numerical error)}.  We compare the performance when varying the number of eigenvalues used. The same prototype part {and cylinder} as in Section \ref{RL} is used.

\begin{table}[H]
\begin{center}
\begin{tabular}{c|c|c|c|c|c}
\hline
 & \multicolumn{5}{c}{Eigenvalues used:} \\
 \cline{2-6}
 & 2nd-5th & 2nd-15th & 2nd-50th & 2nd-75th & 2nd-100th\\
\hline
{In-control cylinder}& 0.0496 & 0.0501 &  0.0527 & 0.0504 & 0.0540\\
\hline
{In-control prototype part}& 0.0478 & 0.0465 & 0.0505 & 0.0473 & 0.0550\\
\hline
\end{tabular}
\caption{In-control alarm probability of the ``Phase I'' scheme by \cite{capizzi2017phase} 
applied to a cylinder and the part design used before, with different number {of LB eigenvalues}. The nominal false alarm probability is $\alpha=0.05$. Results are obtained from 10,000 replications. All parameters are  {at their default values} as in the R package {\bf dfphase1}.}
\label{PhaseI:IC}
\end{center}
\end{table}
Table \ref{PhaseI:IC}  {summarizes the in-control} false alarm probability (FAP), where  {similarly} to Section {\ref{RL}}, we sampled and permuted 50 IC parts from pre-simulated 40,000 IC parts instead of simulating new parts for new replications to reduce  {the} computational  {effort}. As the table shows, all cases have  {a} FAP close to the nominal $\alpha=0.05$, indicating that as long as the FAP is concerned, the  method works well regardless the number of eigenvalues selected.

 {The out-of-control alarm probabilities for the cylindrical parts with increasing ``barrel" shape parametrized by parameter $\delta$ are shown in Table \ref{PhaseI:OC}}.  Each replication consists of 25 simulated IC parts followed by 25 simulated OC parts under the same isotropic noise as before (N$(0, 0.05^2 {\bf I}_3)$). The table indicates how the power to detect changes in the shape of the cylinders is concentrated in the lower part of the spectrum. Using    up to the 100th eigenvalue is counterproductive: the detection capability goes down as the higher eigenvalues are associated with geometrical noise. Using the first 15 eigenvalues, as was also recommended for Phase II, provides good detection power in Phase I as well.

\begin{table}[H]
\begin{center}
\begin{tabular}{c|c|c|c|c|c}
\hline
 & \multicolumn{5}{c}{Eigenvalues used:} \\
 \cline{2-6}
 & 2nd-5th & 2nd-15th & 2nd-50th & 2nd-75th & 2nd-100th\\
\hline
$\delta=0.0005$ & 0.4224 & 0.4860 & 0.2360 & 0.1326 & 0.0553\\
\hline	
$\delta=0.005$ &1 & 1& 1 & 1&0.0741\\
\hline
$\delta=0.5$ & 1& 1& 1& 1&0.0948\\
\hline
$\delta=1$ & 1& 1& 1&1&0.0929\\
\hline
$\delta=10$ &1 &1 &1 &1&0.0876\\
\hline
\end{tabular}
\caption{Out-of-control alarm probability of the ``Phase I'' scheme by \cite{capizzi2017phase} for the cylindrical parts with increasing out-of-control parameter $\delta$. Different number {of LB eigenvalues}  {were} investigated. The IC nominal FAP is $\alpha=0.05$. Results are obtained from 10,000 replications.  All default parameters in R package {\bf dfphase1}  {were used.}}
\label{PhaseI:OC}
\end{center}
\end{table}

 {The OC alarm probabilities for the defective parts displayed in Figure \ref{fig:10} are shown in Table \ref{tab:7}. Each new set of ``Phase 1" data consists of 25 simulated IC parts and 25 simulated OC parts with the same isotropic noise level as before. As it can be seen, the protrusion defect is the easiest to detect, followed by chipped \#2 defect, with the chipped \#1 type of defect being the hardest to detect. This is consistent with our OC results from ``Phase II" (Table \ref{ARL:OC-testpart}). Similarly to the cylindrical-barrel defect parts, using the top 15 eigenvalues provides good detection capabilities, unless the change to detect is small, which is the case of the chipped parts. In such case, using up tp the 75th eigenvalue provides better detection, but again, adding eigenvalues up to the 100th is counterproductive due to their modeling of noise.
 
\begin{table}[H]
\begin{center}
\begin{tabular}{c|c|c|c|c|c}
\hline
{}&\multicolumn{5}{c}{Eigenvalues used:}\\
\cline{2-6}
Defect type& 2nd-5th & 2nd-15th   & 2nd-50th  & 2nd-75th  & 2nd-100th \\
\hline
Chipped \#1	&0.0595 & 0.0640 & 0.0709 & 0.0975 & 0.0531\\
\hline
Chipped \#2	&0.0920 & 0.0969 & 0.1534 & 0.5443 & 0.0562\\
\hline
Protrusion		& 0.9001 &0.9952  &1  &1 & 0.0921 \\
\hline
\end{tabular}
\caption{Out-of-control alarm probabilities of the ``Phase I" SPC scheme by \cite{capizzi2017phase} for the part defects in Figure \ref{fig:10}. Different number were investigated. The nominal in-control FAP is $\alpha=0.05$. Results obtained from 10,000 replications. All default parameters in Rpackage {\bf dfphase1} were used.}\vspace{-0.5cm}
\label{tab:7}
\end{center}
\end{table}

\section{CONCLUSIONS AND FURTHER WORK}

We have presented a fundamentally new approach for the Statistical Process Control of discrete-part manufacturing processes based on intrinsic geometrical properties of the sequence of parts that, contrary to existing methods, does not require registration of the parts and does not require equal number of points per part. Our proposal brings SPC closer to Computer Graphics/Vision methods. The SPC problem, however, is inherently different than the shape similarity problem from these other fields, given that contrary to them, a method to be useful for SPC must be able to distinguish small but significant shape and size differences in a sequence of very similar parts measured with noise, avoiding false alarms but considering increments in noise a potential additional source of an out-of-control condition. In contrast, computer graphics/vision methods for shape similarity assessment typically aim to detect large shape differences in a manner that is robust with respect to any measurement noise, which (if existent) is filtered out, and usually neglect differences in size.

The main differential-geometric tool we use is the unnormalized  spectrum of the discrete Laplace-Beltrami operator, cropped to consider only its lower part.  We discussed two different discrete LB operator approximations which are symmetrizable (ensuring a real spectrum and providing computational advantages) and pointwise convergent (providing theoretical guarantees), and adopted the localized mesh Laplacian of \cite{Li2015} due to its sparseness. Other discrete approximations of the LB operator are also symmetrizable, based on Finite Element Methods (FEM, \citealt{ReuterShapeDNA}), and we leave their study and comparison with the Localized Mesh Laplacian used here for future work. The LB-spectrum chart method is intrinsic and hence avoids registration of the parts, which is a hard-to-solve combinatorial problem. 

Given the non-normality of the discrete LB spectrum, we proposed to use (with some  modifications) a multivariate, nonparametric permutation-based control chart due to \cite{Chen2016} for on-line or ``Phase II" SPC and a similar permutation and rank-based approach for the startup or "Phase I"  due to \cite{capizzi2017phase}. Run length analyses and detection probability assessments, respectively, indicate the practical feasibility of the methods, even with relative large meshes (with tens of thousands of points) on a modest {desktop} computer {assuming enough storage}. The on-line (Phase II) method is especially sensitive to detect small changes in the shape or size of the surfaces, while providing an easy to tune in-control Average Run Length. The Phase I method {applied to only the first several eigenvalues has} excellent detection performance while controlling the false alarm probability.
We compared our Phase II method with a nonparametric univariate chart based on registration of the parts using the Iterated Closest Point (ICP) algorithm, considering its objective function as a monitoring statistic, but found the LB-spectrum chart to be much more sensitive to detect process changes. An ICP-based method was presented to determine the localization of the defect on the part surface as a post-alarm diagnostic only to be used after the generic, or overall SPC mechanism provided by the LB spectrum chart, triggers an alarm. Phase II run length performance comparisons were made also for isotropic and non-isotropic noise, and further comparisons with a registration-based SPC method shows very competitive behavior for our method, even under conditions that clearly favor registration methods (small, equal size lattice meshes).


We focused in this paper on surface data, where the intrinsic dimension of the manifold is 2 and the topology of the object has genus 0. Our methods carry over to the case of voxel data, where the intrinsic dimension of the manifold is 3 and objects with holes and internal features can be modeled. These are of particular interest in SPC of additive manufacturing data obtained via computed tomography scans of a part, in order to determine the inner features of printed parts. The intrinsically 3-manifold data can be represented with a tetrahedralization, and FEM methods exist for approximating the LB operator from such data structure \citep{ReuterVoxel}, thus we will consider voxel extensions of our methods in future work. 

Further work is also needed to develop charts that detect not only changes in the mean geometry of a part, but also changes in the overall variance of noise. Our methods assumed no systematic local bias due to optical aberration in the scanner, and extensions to deal with such bias, if significant, are of interest if calibration is not efficient. 

{\bf Acknowledgements.-}  We thank four anonymous referees, an Associate Editor, and the Editor (Prof. Roshan Joseph) for several comments and suggestions which have resulted in an improved manuscript.
We also thank Professors Bianca Colosimo (Politecnico di Milano, Italy) and Massimo Pacella (Universit\`a del Salento, Italy) for their kindness in making available the data and code from their paper \citep{colosimo2014profile}.\\

\bigskip
\begin{center}
{\large\bf SUPPLEMENTARY MATERIALS}
\end{center}

\begin{description}
\item[Appendices]: A) Proofs; B) Exact moments of the DFEWMA chart statistics; C) Further discussion on the heat equation and the LB operator of manifolds and graphs; D) Other intrinsic geometrical statistics and their use for SPC.
\item[Matlab code] for the computation of the Localized Mesh Laplacian and for the modified DFEWMA control chart in the examples.  
\item[Data set:] Prototypical part CAD model and CAD model for cylinder (mesh data), both in-control noise-free and noise-free defect versions included.
\end{description}
\baselineskip=12pt
\bibliographystyle{authordate1}
\bibliography{bibliography.bib}\vspace{-0.5cm}
\newpage

\begin{appendix}
\baselineskip=16pt
\section{Appendix. Proofs.}
\label{App2}

\subsection{Proof of Lemma 1. Mesh Laplacian derivation}
We show how the discrete Mesh Laplacian can be written as the matrix $L_k^t=D-W$. Define $\mc K$=given triangulation, $T_{\mc K}$=set of all triangles in $\mc K$, $V_{\mc K}$=set of all vertices in $\mc K$, $A(T), V(T)$=area and set of vertices of triangle $T$, {respectively.} The discrete Mesh Laplacian can be written as follows:
\[\begin{aligned}
L^t_{\mc K}f(p_i) &= \frac{1}{4\pi t^2}\sum_{T\in T_{\mc K}}\frac{A(T)}{3}\sum_{p_j\in V(T)}e^{-||p_i-p_j||^2/4t}(f(p_i)-f(p_j))\\
&=\frac{1}{4\pi t^2}\sum_{T\in T_{\mc K}}\sum_{p_j \in V(T)}\frac{A(T)}{3}e^{-||p_i-p_j||^2/4t}(f(p_i)-f(p_j)) \\
&\qquad (\text{For each triangle, look at all its vertices})\\
&=\frac{1}{4\pi t^2}\sum_{p_j\in V_{\mc K}}\sum_{T: p_j\in V(T)}\frac{A(T)}{3}e^{-||p_i-p_j||^2/4t}(f(p_i)-f(p_j)) \\
&\qquad (\text{For each vertex, look at all triangles associated with it})\\
&=\frac{1}{12\pi t^2}\sum_{p_j\in V_{\mc K}}\sum_{T: p_j\in V(T)}A(T)e^{-||p_i-p_j||^2/4t}(f(p_i)-f(p_j)) \\
&=\left(\sum_{p_j\in V_{\mc K}} \frac{1}{12\pi t^2}\sum_{T: p_j\in V(T)}A(T)e^{-||p_i-p_j||^2/4t}\right) f(p_i) \\
&\qquad\qquad - \sum_{p_j\in V_{\mc K}}\left(\frac{1}{12\pi t^2}\sum_{T: p_j\in V(T)}A(T)e^{-||p_i-p_j||^2/4t}\right) f(p_j)
\end{aligned}\]
{For any function $f$,  we first discretize it} based on the points in the mesh as a vector $f=\begin{pmatrix} f(p_1) & f(p_2) & \cdots & f(p_m)\end{pmatrix}^T$. Then let $W_{ij}=\frac{1}{12\pi t^2}\sum_{T: p_j\in V(T)}A(T)e^{-||p_i-p_j||^2/4t}$ and $D_{ii}=\sum_j W_{ij}$, we have\vspace{-0.2cm}
\[\begin{aligned} L^t_{\mc K} f(p_i) &= \left[ D_{ii} f(p_i) - \begin{pmatrix}
    W_{i1} & W_{i2} & \cdots & W_{im}\end{pmatrix}
    \begin{pmatrix} f(p_1) \\ f(p_2) \\ \vdots \\ f(p_m)
     \end{pmatrix} \right] \\
     & = \left[ \begin{pmatrix} 0 & 0 & \cdots & D_{ii} & \cdots & 0\end{pmatrix} 
      \begin{pmatrix} f(p_1) \\ f(p_2) \\ \vdots \\ f(p_i)\\ \vdots \\ f(p_m) \end{pmatrix}  - \begin{pmatrix}
    W_{i1} & W_{i2} & \cdots & W_{im}\end{pmatrix}
    \begin{pmatrix} f(p_1) \\ f(p_2) \\ \vdots \\ f(p_m)
     \end{pmatrix} \right]    
\end{aligned}\]
Define $D= \mbox{diag}[D_{ii}]$ and $W=[W_{ij}]$, then $L^t_{\mc K} f=(D-W)f$ and $L^t_K=D-W$. This is in agreement with the Graph Laplacian, defined as $L=D-W$ where the weights $W$ are given by the adjacency matrix of the graph. $ \; \; \blacksquare$

\subsection{Proof of Lemma 2, convergence of the spectrum of the Localized Mesh Laplacian.}

Here we prove the convergence of the spectrum of the Localized Mesh Laplacian (\ref{Localized}) as the mesh $\mc K$ gets finer and  closely follows the manifold {$\mc M$}, a result that in itself is new but follows directly from previous results in \cite{Belkin2008Discrete}, \cite{Li2015} and \cite{dey2010convergence}. Following these authors, denote by $\rho(\mc M)$ the {\em reach} of the manifold, defined as the infimum of the closest distance from any point $p \in \mc M$ and the medial axis of $\mc M$. Define an $({\epsilon,\eta})$ approximation mesh to a manifold $\mc M$ as a triangulation $\mc K$ where: 1) for any point $p_i \in \mc M$, there is vertex in $\mc K$ that is at most $\epsilon \rho(\mc M)$ away, and 2)  
for a face $t \in \mc K$ and vertex $p \in t$, the angle between the normals at $p$ and $t$ is at most $\eta$, where $\eta = O(\epsilon)$. Condition one assures the mesh is fine enough and condition 2 ensures the distortion between $\mc K$ and $\mc M$ is small. 

Consider an $(\epsilon, \eta)$ mesh approximation $\mc K$ to a smooth surface $\mc M \subset \mathbb{R}^3$.
{In \cite{dey2010convergence}, an intermediate operator $\mc L_{t}^\mc M$ defined as
$$\mc L_{t}^\mc M f(x)=\frac{1}{4\pi t^2}\int_\mc M e^{-\frac{||x-y||^2}{4t}}(f(x)-f(y))dy$$
is used to connect the Mesh Laplacian $L^t_{\mc K}$ in (\ref{MeshLB}) and the manifold LB operator $\Delta_\mc M$. The authors proved the pointwise convergence between $\mc L_{t}^\mc M f$ and $\Delta_\mc Mf$, and between $L^t_{\mc K}f$ and $\mc L_{t}^\mc M f$ in sequence, and finally arrived at the convergence of the spectrum of $L^t_{\mc K}$ to the true LB spectrum as $\epsilon \rightarrow 0, t \rightarrow 0$, and $\epsilon/t^4 \rightarrow 0$. 

As \cite{dey2010convergence} indicate, to also show convergence between the corresponding spectra of $L_{\mc K}^t$ and $\Delta_\mc M$, a stronger (than pointwise) convergence is needed between the intermediate operator $\mc L_{t}^\mc M$  and the Mesh Laplacian $L^t_{\mc K}$. These authors show convergence between these operators in a Sobolev 2-norm, defined for a function $f \in H_2$ (Sobolev space) by $
||f||_{H_2} = \sqrt{ ||f^{(1)}+f^{(2)}||^2}$. Given the small differences between the Mesh Laplacian (\ref{MeshLB}) and the Localized Mesh Laplacian (\ref{Localized}), we follow their proof using the latter Laplacian, which requires geodesic distances on $\mc M$.

First, we modify the intermediate operator $\mc L_{t}^\mc M$ to use the geodesic distances instead of the Euclidean distances in the exponential term. Theorem 1 in \cite{Li2015} proved the pointwise convergence between the new intermediate operator and the true LB operator as $t\to 0$. To show convergence in $H_2$ between the Localized Mesh Laplacian and the new intermediate operator and complete the proof, we need to substitute the geodesic distance for the Euclidean distance in the heat kernel $G_t(x,y)\triangleq e^{-\frac{[\text{dist}(x,y)]^2}{4t}}$, whose structure is explicitly used only in  Lemma 4.3 in \cite{dey2010convergence}, which asserts that the derivatives needed in the $H_2$ norm have the form:
$$\frac{\partial^i G_t(x,y)}{\partial x^i}=\sum_{j=0}^{\lfloor\frac{i}{2}\rfloor}O(i^i)\frac{[\text{dist}(x,y)]^{i-2j}}{(2t)^{i-j}}G_t(x,y)$$
\cite{dey2010convergence} proved this Lemma for $\text{dist}(x,y)=||x-y||$. We need to prove it for $\text{dist}(x,y)=g(x,y)$, the geodesic distance between $x$ and $y$, which can be done in 2 steps, as follows.\\
\\
First we observe that
$$\frac{\partial}{\partial x}e^{-\frac{||x-y||^2}{4t}}= -\frac{||x-y||}{2t}e^{-\frac{||x-y||^2}{4t}}\frac{\partial ||x-y||}{\partial x}, \quad \frac{\partial}{\partial x}e^{-\frac{g(x,y)^2}{4t}}= -\frac{g(x,y)}{2t}e^{-\frac{g(x,y)^2}{4t}}\frac{\partial g(x,y)}{\partial x}$$
where the term $\frac{[\text{dist}(x,y)]^{i-2j}}{(2t)^{i-j}}G_t(x,y)$ has exactly the same format (and also the same power) regardless the distance used. This equivalence remains true for higher order derivatives, because this structure (and also the power in $O(i^i)$) comes from differentiating an exponential term (namely $G_t(x,y)$) repeatedly, regardless of which distance is used.

Secondly, we consider the difference between $\frac{\partial ||x-y||}{\partial x}$ and $\frac{\partial g(x,y)}{\partial x}$. Since small neighborhoods in Riemannian manifolds can be seen as Euclidean spaces, $\frac{\partial ||x-y||}{\partial x}$ and $\frac{\partial g(x,y)}{\partial x}$ at least have the same magnitude, and the exact ratio can be omitted in the $O(i^i)$ term. Putting these facts together shows how Lemma 4.3 also holds for the geodesic distance as well. Once this is done, all remaining steps in the proof of \cite{dey2010convergence} can be exactly followed to obtain the convergence of the spectrum of the Localized Mesh Laplacian to the true LB spectrum under the same limit conditions, namely, $\epsilon \rightarrow 0, t \rightarrow 0$, and $\epsilon/t^4 \rightarrow 0$. $\; \; \blacksquare$
}

\subsection{Proof of Proposition 1, on the real spectrum of the LB operator}

Here we prove the following:

{\bf Proposition 1.} The Mesh Laplacian (\ref{MeshLB}) and the Localized Mesh Laplacian (\ref{Localized}) can be written as (\ref{symmetrizable}) and therefore their eigenvalues are all real.

{\em Proof.} {From {Lemma} 1, both the Mesh Laplacian and the Localized Mesh Laplacian can be written  {as}
$
L^t_{\mc K}=D-W
$, 
where $W_{ij}=\frac{1}{12\pi t^2} A(p_i, p_j) e^{-dist(p_i, p_j)^2/4t}$ and $D_{jj}= \sum_i W_{ij}$ is diagonal, for $i=1,...,m$ and $j=1,...,m$. The only differences between these two  {discrete} Laplacians are the definitions of area $A(p_i, p_j)$ and distance $dist(p_i, p_j)$, which  {are} summarized in the following table:
\begin{center}
\begin{tabular}{c|c|c}
\hline
	& {Mesh Laplacian} & {Localized Mesh Laplacian} \\ 
\hline 
$A(p_i, p_j)$ & $A(p_j)$  & $\begin{cases} A(p_j) &\text{if }dist(p_i, p_j)\leq r \\ 0 & \text{if }dist(p_i, p_j)> r \end{cases}$ \\
\hline
$dist(p_i, p_j)$ & Euclidean distance & Geodesic distance \\
\hline
\end{tabular}
\end{center}
Where  {recall $A(p_j)$} is the area of the one ring neighborhood of point $p_j$. Now define a diagonal matrix $B$ where $B_{ii}=A(p_i)$, which is obviously positive definite.  {Pre-multiplying a matrix by} $B$ is the same as multiplying the $i$th row of the matrix  {times} the $i$th diagonal element in $B$, $i=1,...,m$, so matrix $BW$ has elements: 
\[
(BW)_{ij}=A(p_i)W_{ij}=\frac{1}{12 \pi t^2} A(p_i)A(p_i, p_j) e^{-dist(p_i, p_j)^2/4t}
\]
It is easy to see that $A(p_i)A(p_i, p_j)=A(p_j)A(p_j, p_i)$ holds with either of the two area definitions above, and indices $i$ and $j$ are interchangeable in  {both types of} distances. Therefore, $BW$ is symmetric for both  {types of discrete} Laplacians. 
 {In addition,} $BD$ is a diagonal matrix and automatically symmetric. So $BL^t_{\mc K}=BD-BW {\triangleq} L$ is symmetric, and $L^t_{\mc K}=B^{-1}L$ is the multiplication of two symmetric, positive semi-definite ($B$ is positive definite) matrices.} Therefore, both discrete Laplacians, (\ref{MeshLB}) and (\ref{Localized}) are symmetrizable and thus always have real eigenvalues. $\blacksquare$\\

\section{Appendix. Exact moments for DFEWMA chart statistic}
\label{App3}
Here we give the derivation of the correct moments {in} the statistic (\ref{DFEWMA}). Since this is a weighted sum,  the {\em covariances} between the terms added need to be considered\footnote{We point out that the expression for $T_{jn}(w,\lambda)$ in \cite{Chen2016} has an extra $w$ in the numerator (the mean), and the denominator (the variance) is only correct for $\lambda=0$.}.


{First consider $R_{jni}$ as defined in Section \ref{sec:4.2}.} Under the null hypothesis, all parts come from the same distribution. Thus, $R_{jni}$ can be any number from 1 to $m_0+n$ with equal probability, {and therefore, it is easy to show that} 
$$E[R_{jni}]=\frac{m_0+n+1}{2}, \quad \mbox{and } \quad \mbox{Var}[R_{jni}]=\frac{(m_0+n+1)(m_0+n-1)}{12}.$$
\\
\baselineskip=12pt
{Now consider $\sum_{i=n-w+1}^n (1-\lambda)^{n-i} R_{jni}$. Its expectation is: }
\begin{eqnarray*}
E\left[\sum_{i=n-w+1}^n (1-\lambda)^{n-i} R_{jni}\right]&=&\sum_{i=n-w+1}^n (1-\lambda)^{n-i} E[R_{jni}]=\sum_{i=n-w+1}^n (1-\lambda)^{n-i} \frac{m_0+n+1}{2}\end{eqnarray*}
When $\lambda=0$, $$E\left[\sum_{i=n-w+1}^n (1-\lambda)^{n-i} R_{jni}\right]=\sum_{i=n-w+1}^n E[R_{jni}]=\frac{w(m_0+1+n)}{2}$$ 
When $\lambda\neq0$, $$E\left[\sum_{i=n-w+1}^n (1-\lambda)^{n-i} R_{jni}\right]=\frac{1-(1-\lambda)^w}{\lambda}\frac{m_0+n+1}{2}\triangleq L_1(\lambda,w)\frac{m_0+n+1}{2}$$ where $\lim_{\lambda \rightarrow 0}\; L_1(\lambda,w) = w$, i.e. 
$$\lim_{\lambda \rightarrow 0} E\left[\sum_{i=n-w+1}^n (1-\lambda)^{n-i} R_{jni}\right] = E\left[\sum_{i=n-w+1}^n R_{jni}\right]$$

\nd Next we look at the variance:
\begin{eqnarray*}
Var\left(\sum_{i=n-w+1}^n (1-\lambda)^{n-i} R_{jni}\right)&=&\sum_{i=n-w+1}^n (1-\lambda)^{2(n-i)} Var(R_{jni}) \\
&+& 2\sum_{n-w+1\leq i<k\leq n}Cov((1-\lambda)^{n-i}R_{jni},(1-\lambda)^{n-k}R_{jnk})\\
&=&\sum_{i=n-w+1}^n (1-\lambda)^{2(n-i)} Var(R_{jni}) \\
&+& 2\sum_{n-w+1\leq i<k\leq n} (1-\lambda)^{2n-i-k}Cov(R_{jni},R_{jnk})
\end{eqnarray*}

In particular, let $N=m_0+n$ and consider:
\begin{eqnarray*}
E[R_{jni}R_{jnk}]&=&\sum_{a=1}^{N-1}\sum_{b=a+1}^N ab\frac{1}{N(N-1)/2}\\
&=&\frac{2}{N(N-1)}\sum_{a=1}^{N-1}a\sum_{b=a+1}^N b \\
&=&\frac{2}{N(N-1)}\sum_{a=1}^{N-1}a(a+N+1)(N-a)/2 \\
&=&\frac{1}{N(N-1)}\sum_{a=1}^{N-1}[-a^3-a^2+a(N^2+N)] \\
\end{eqnarray*}
\begin{eqnarray*}
&=&\frac{1}{N(N-1)}\frac{N(N^2-1)(3N+2)}{12} \\
&=&\frac{(N+1)(3N+2)}{12} \\
\end{eqnarray*}
\begin{eqnarray*}
Cov(R_{jni},R_{jnk})&=&E[R_{jni}R_{jnk}]-E[R_{jni}]E[R_{jnk}]\\
&=&\frac{(N+1)(3N+2)}{12} - \left(\frac{N+1}{2}\right)^2 \\
&=&-\frac{N+1}{12} = -\frac{m_0+n+1}{12} 
\end{eqnarray*}

When $\lambda=0$, the variance reduces to:
$$\begin{aligned}
Var\left(\sum_{i=n-w+1}^n (1-\lambda)^{n-i} R_{jni}\right)&=\sum_{i=n-w+1}^n (1-\lambda)^{2(n-i)} Var(R_{jni}) \\
&\qquad +2\sum_{n-w+1\leq i<k\leq n} (1-\lambda)^{2n-i-k}Cov(R_{jni},R_{jnk})\\
&=w\frac{(m_0+n+1)(m_0+n-1)}{12} -2\frac{w(w-1)}{2} \frac{m_0+n+1}{12}\\
&=\frac{w(m_0+n+1)(m_0+n-w)}{12}
\end{aligned}$$
When $\lambda\neq0$,
\begin{eqnarray*}
Var\left(\sum_{i=n-w+1}^n (1-\lambda)^{n-i} R_{jni}\right)&=&\sum_{i=n-w+1}^n (1-\lambda)^{2(n-i)} Var(R_{jni}) \\
&\qquad& +2\sum_{n-w+1\leq i<k\leq n} (1-\lambda)^{2n-i-k}Cov(R_{jni},R_{jnk})\\
&=&\left(\sum_{i=n-w+1}^n (1-\lambda)^{2(n-i)}\right)\frac{(m_0+n+1)(m_0+n-1)}{12} \\
&\qquad& -\left(\sum_{i=n-w+1}^{n-1}\sum_{k=i+1}^n (1-\lambda)^{2n-i-k}\right) \frac{m_0+n+1}{6}\\
&=&\frac{1-(1-\lambda)^{2w}}{2\lambda-\lambda^2} \frac{(m_0+n+1)(m_0+n-1)}{12} \\
&\qquad& -\left(\frac{1-\lambda-(1-\lambda)^w}{\lambda^2} -\frac{(1-\lambda)^2-(1-\lambda)^{2w}}{\lambda^2(2-\lambda)}\right) \frac{m_0+n+1}{6}\\
&\triangleq& L_2(\lambda,w)  \frac{(m_0+n+1)(m_0+n-1)}{12} + L_3(\lambda,w)\frac{m_0+n+1}{6} 
\end{eqnarray*}
where $\lim_{\lambda \rightarrow 0} L_2(\lambda,w) = w$ and $\lim_{\lambda \rightarrow 0} L_3(\lambda,w) = \frac{w(w-1)}{2}$, i.e. 
$$\lim_{\lambda \rightarrow 0} Var\left(\sum_{i=n-w+1}^n (1-\lambda)^{n-i} R_{jni}\right) =  \frac{w \left( m+n+1 \right)  \left( m_0+n-w \right)}{12}=Var\left(\sum_{i=n-w+1}^n R_{jni}\right)$$
\\
Summarizing, the expectation equals to:
\be
E\left[\sum_{i=n-w+1}^n (1-\lambda)^{n-i} R_{jni}\right] = \begin{cases}
\frac{w(m_0+n+1)}{2} & \lambda=0 \\
\frac{1-(1-\lambda)^w}{\lambda}\frac{m_0+n+1}{2} & \lambda\neq0 \end{cases}
\ee
and the variance equals to:
\begin{small}
 {$$Var\left(\sum_{i=n-w+1}^n (1-\lambda)^{n-i} R_{jni}\right)=\begin{cases}
\frac{w(m_0+n+1)(m_0+n-w)}{12} & \lambda=0 \\
\frac{1-(1-\lambda)^{2w}}{2\lambda-\lambda^2} \frac{(m_0+n+1)(m_0+n-1)}{12} \; -\; \\
\quad \left(\frac{1-\lambda-(1-\lambda)^w}{\lambda^2} -\frac{(1-\lambda)^2-(1-\lambda)^{2w}}{\lambda^2(2-\lambda)}\right) \frac{m_0+n+1}{6} & \lambda\neq0
\end{cases}$$}
\end{small}
\baselineskip=16pt
\section{Appendix: further discussion on the LB operator of manifolds and graphs.}
\label{AppD}
\subsection{The heat equation and the LB operator}

 The LB operator emerges from the spatial part of the solution to the heat and wave partial differential equations on a manifold $\mc{M}$ \citep{Evans}. The intuition is that to model both heat and wave phenomena on a manifold it is necessary to consider the geometry or curvature of $\mc{M}$, and the LB operator encodes it. A heat diffusion process over $\mathcal{M}$, for our purposes, the surface of a 3D object, is governed by the {\em heat equation} (or diffusion equation):
\be
\Delta_{\mathcal{M}} u(x,t) = \frac{\partial u(x,t)}{\partial t}
\label{HeatEq}
\ee
where $\Delta_{\mc{M}}$ is the LB operator on $\mc{M}$ and $u(x,t)$ is the temperature at $x$ at time $t$, assumed enough times differentiable in each argument. The heat equation intuitively says that the rate of change in time of the temperature is proportional to the ``curvature" of $u$ in $\mc{M}$. 

Separation of variables $u(x,t)=f(x)g(t)$ in the heat equation leads to $f(x) g'(t) = \Delta f(x) g(t)$, which, dividing both sides by $f(x)g(t)$ we get $\frac{g'(t)}{g(t)} = \frac{\Delta f(x)}{f(x)} = \lambda$ from which we obtain the so-called Helmholtz differential equation (\ref{Helmholtz}):
\[
\Delta f(x) = \lambda f(x)
\]
{Similarly, the Helmholtz equation can also originate from the spatial part of the solution to the wave equation $ \Delta u(x,t)= \frac{\partial^2 u(x,t)}{\partial t^2}$}. 

Given an initial heat distribution $u(x,0)= f :\mc{M} \rightarrow \mathbb{R}$, the complete or fundamental space-time solution to the heat equation is given by the integral equation \citep[pg. 46]{Evans}:
\[
u(x,t) = H_t f(x) = \int_{\mc{M}} k_t(x,y) f(y) dy
\]
where the bilinear function $k_t(x,y): \mathbb{R}^+ \times \mc{M} \times \mc{M} \rightarrow \mathbb{R}$ is called the {\em heat kernel} which can be thought of as the amount of heat transmitted from $x$ to $y$ in time $t$ if {initially} there is a unit heat source at $x$ and where $dy$ stands for area (of $\mc{M}$).  $H_t$ is called the {\em heat operator}.  If $\mc M$ has boundaries, the solution to the heat and wave equations requires additional conditions such as the so-called Dirichlet boundary condition $u(x,t)=0 $ for all $x \in \partial \mc M$(so the boundary $\partial \mc M$ acts as an absolute refrigerator for all $t>0$ in the case of the heat equation).

In case of a flat 2D space (i.e., when $\mc{M} = \mathbb{R}^2$), the heat kernel is given by the Gaussian function \citep{Belkin2008}:
\[
k_t(x,y) = \frac{1}{4 \pi t}e^{-\frac{||x-y||^2}{4t}}.
\]
(in the more general manifold case, when $\mc M \subset \mathbb{R}^n$, the heat kernel can be obtained only after obtaining the spectrum of the LB operator, using expression (\ref{heatKernel}) shown in Appendix \ref{sec:7}). Note how the Mesh Laplacian (\ref{MeshLB}) is based on the heat kernel.

%
%
%

If $\mc M = \mathbb R^2$, the eigenfunctions $f(u,v)_i$  satisfying the Helmholtz differential equation ($\Delta_{\mc M} f = \lambda f$) can be interpreted as the modes of vibration of a 2D drum membrane with resonant or natural frequencies $\lambda_i$. The question of whether one can determine the shape of the drum from its spectrum entertained mathematical physicists for many years until it was shown by  \cite{Gordon} that there are pairs of 2D drums that have the same spectrum, yet their shapes are different.  In general, for $\mc M \subset \mathbb R^3$, this result implies that the geometric information contained in the spectrum is not enough to  {uniquely identify} the shape of the surface $\mc M$. However, these pairs of figures are always concave polygons and correspond segment by segment, so they are rare. 

\subsection{Relation between the mesh Laplacians and the combinatorial graph Laplacian} 

The discrete Mesh Laplacian  in equation (\ref{MeshLB}) and the Localized Mesh Laplacian (\ref{Localized}), have the same form as the {\em combinatorial Graph Laplacian} \citep{Chung},  {$L=D-A_{\mc K}$, where $A_{\mc K}$ is the adjacency matrix  {of the mesh} $\mc K$} and $D$ is a diagonal matrix with entries equal to the degree of each vertex. The graph Laplacian is an operator applied to a function defined on the vertices of the graph. It corresponds to the discrete Laplacian if the edge weights are $W_{ij}=1$, so the discrete LB operator (\ref{MeshLB}) can be thought of as a weighted version of the combinatorial Laplacian where the edge weights are given by the integrated heat kernel (with either euclidean or geodesic distances over the neighborhood of a point, depending on the approximation used) over one third of the {area of the one-ring neighborhood of either of its two end points.} The graph Laplacian arises when modeling a diffusion process on a network or mesh. Let $\psi_i$ be the amount of some substance at vertex $i$, and $c(\psi_j-\psi_i)$ be the rate at which the substance flows from vertex $j$ to vertex $i$, where $c$ is a constant. Then 
\[
\frac{d \psi_i}{d t} = c \sum_j A_{ij}(\psi_j-\psi_i)
\]
where $A_{ij}$ is the $(i,j)$th element of $A_\mc{K}$. It is easy to show \citep{Newman} that
\[
\frac{\partial \psi}{\partial t} = c(A_\mc{K}-D) \psi = -c L  \psi
\]
where $\psi$ is a vector with all $\psi_i$'s, from which the diffusion equation on a network results, compare to the heat equation (\ref{HeatEq}).

The spectrum of the combinatorial graph Laplacian $L$ has some properties with counterparts in the approximated spectrum of $L_{\mc K}^t$ as the latter is also based on a mesh or network: first, $\lambda_1=0$, and the algebraic multiplicity of this eigenvalue gives the number of connected components in the graph. The first eigenvector is constant, and the second eigenvalue $\lambda_2$ is greater than zero only if the graph is connected ({so} $\lambda_1$ is not repeated). The signs in the second eigenvector, called {\em Fiedler's vector}, can be used to cluster the vertices in two sets, a notion related to the nodal sets of the eigenvector{s} of the LB operator (see \citealt{Chung}).

\subsection{Computation of the true LB operator for a parametric surface}
The following material can be found in Differential Geometry books (e.g. \citealt{Kreyszig}). 
%
%
Here we illustrate with an example the first fundamental form and the LB operator for a parametric surface ($k=2$). In this case, (see Definition 3) the first fundamental form is given by the quadratic form:
\[
ds^2 = g_{11} (dx^1)^2 + 2 g_{12} dx^1 dx^2 + g_{22} (dx^2)^2
\]
Consider a Torus in $\mathbb{R}^3$  described parametrically by:
\[
{\bf x}(u,v) = ((R+r \cos v) \cos u, (R+r \cos v) \sin u, r \sin v)
\]
where the surface coordinates are $x^1=u$ and $x^2=v$ (see Figure \ref{fig:TorusColored}) and where $R$ is the outer radius and $r$ is the inner radius. Then, $g_{11}=(R+r \cos v)^2$, $g_{12}=0$ and $g_{22} = r^2$, and the first fundamental form is:
\[
ds^2 = (R+r \cos v)^2 (dx^1)^2 + r^2 (dx^2)^2
\]
\begin{figure}[h!]
\centering
\begin{tabular}{ccc}
\includegraphics[scale=0.45]{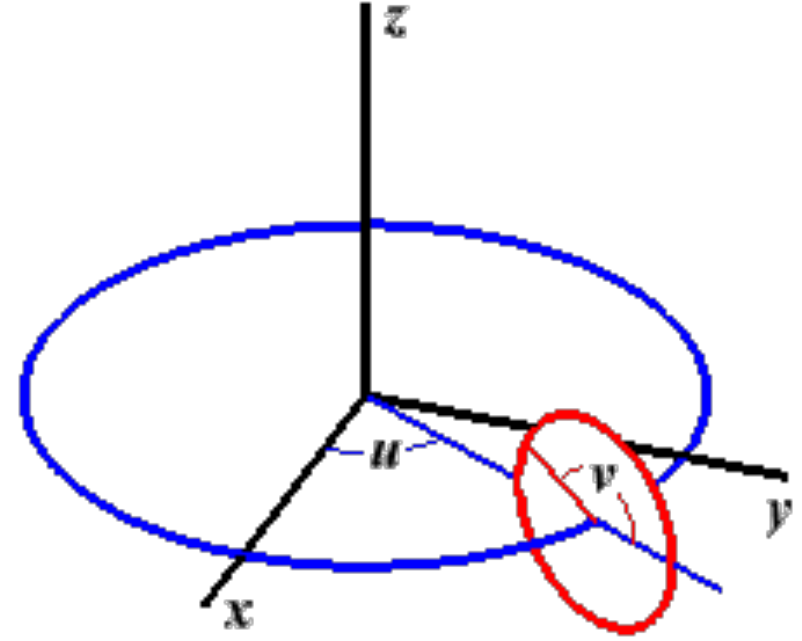}
&\includegraphics[width=6.0cm,height=3.0cm]{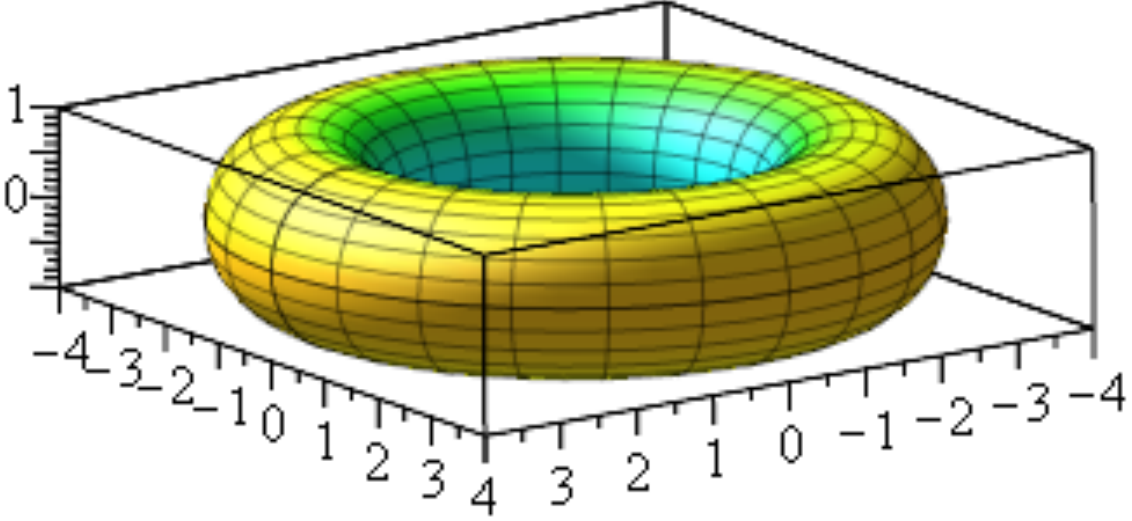}&\includegraphics[width=6.0cm,height=3.0cm]{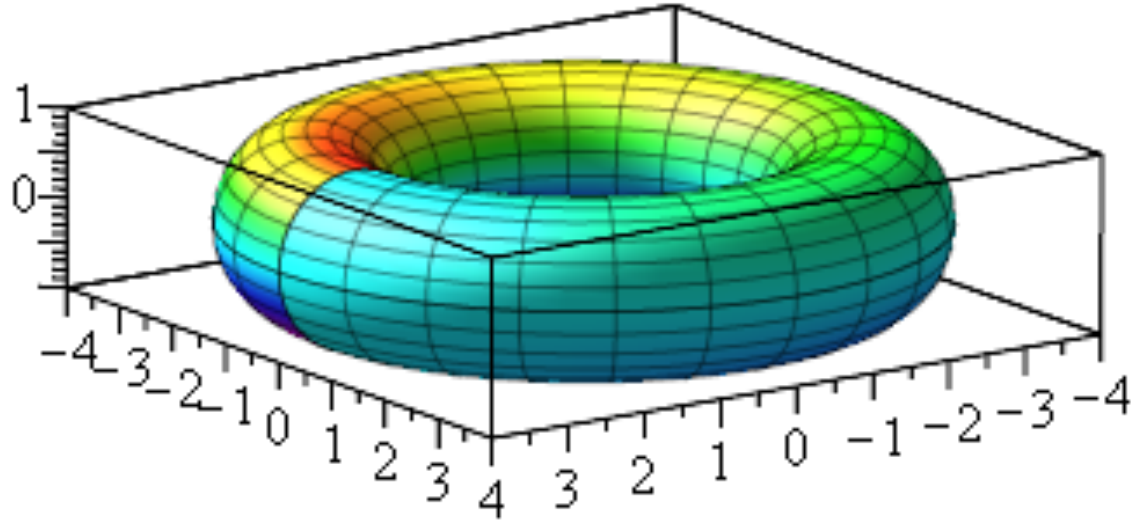}\\
\end{tabular}
\caption{From left to right: Usual torus parametrization, and torus ($R=3,r=1$) colored according to $\Delta_{\mc{M}} f$ for $f=u^2$ and $f=u \cdot v$. Red and blue represent highest to lowest function values respectively. The LB operator on $f=u^2$ is not a function of $u$, the ``horizontal" angle, so note how it is constant for each $u$, varying only transversally to the ``tube" as a function of $v$. For $f=u \cdot v$, in contrast, the Laplacian is a function of both $u$ and $v$.}
\label{fig:TorusColored}
\end{figure}

\nd{Therefore,
$$g=\begin{pmatrix} (R+r \cos v)^2 & 0 \\ 0 & r^2\end{pmatrix} \qquad g^{-1}=\begin{pmatrix} \frac{1}{(R+r \cos v)^2} & 0 \\ 0 & \frac{1}{r^2}\end{pmatrix}$$
From equation (\ref{LBoperator}),
$$\begin{aligned}
\Delta_{\mc{M}} f &= {-}\frac{1}{\sqrt{det(g)}}\sum_{j=1}^k \frac{\partial}{\partial x^j} \left( \sqrt{det(g)} \sum_{i=1}^k g^{ij} \frac{\partial f}{\partial x^i} \right)\\
&={-}\frac{1}{(R+r \cos v)r}\left[\frac{\partial}{\partial u}\left((R+r \cos v)r\frac{1}{(R+r \cos v)^2}\frac{\partial f}{\partial u}\right)+\frac{\partial}{\partial v}\left((R+r \cos v)r\frac{1}{r^2}\frac{\partial f}{\partial v}\right)\right]\\
&={-}\frac{1}{(R+r \cos v)r}\left[\frac{\partial}{\partial u}\left(\frac{r}{R+r \cos v}\frac{\partial f}{\partial u}\right)+\frac{\partial}{\partial v}\left(\frac{R+r \cos v}{r}\frac{\partial f}{\partial v}\right)\right]
\end{aligned}$$}
\nd Consider the function $f=u^2$ defined on the Torus $\mc{M}$. The LB operator applied on $f$ equals
\[
\Delta_{\mc{M}}f =- \frac{2}{(R+r \cos v)^2}
\]
which is only a function of $v$ (Figure \ref{fig:TorusColored}, middle). If $f=u \cdot v$ instead, its Laplacian is:
\[
\Delta_{\mc{M}}f =\frac{u \sin v}{(R+r \cos v)r}
\]
which varies as a function of both angles (Figure \ref{fig:TorusColored}, right).

\section{Appendix: Other intrinsic geometrical statistics and their use for SPC.}
\label{sec:7}

In addition to the spectrum of the LB operator, there are several other descriptors of shape used for object recognition in Computer Vision. Here we review some of these alternatives and report on their applicability to the SPC problem. We focus only on {\em intrinsic} geometrical properties, which can be computed for each object without reference to the ambient space, and hence in principle could be used for SPC schemes that do not require registration. We also consider these statistics as further diagnostics in case an alarm has been triggered, similar to the ICP diagnostic presented in Section \ref{sec:5}.

\subsection{Heat kernel and diffusion distances}

For a long time, researchers in computer vision have tried to find shape descriptors that use the properties of a cloud of points from the point of view of each point in the cloud. In a highly cited paper, \cite{Belongie2002} introduce the concept of a {\em shape context}, a local description of the shape in the vicinity of a given point, and use it for registration of objects. The main idea is to measure the frequency and location of other points in the
neighborhood of each point of a shape and use {differences between} these measures as costs to be
minimized in a classical weighted matching problem (thus points with similar local information tend to be matched), solvable via
Linear Programming. For a point $i$ in an object, {\cite{Belongie2002}}
propose to compute a 2-dimensional histogram where the number of
points nearby are counted. If $r$ is the Euclidean distance between
two points of the shape, the 2-dimensional histogram $h_i(\cdot,\cdot)$ extends along
$log \; r$ and $\theta$, measuring the distance and direction where the
nearby points are located. The 2-dimensional histogram formed by the frequencies
$h_i(\cdot,\cdot)$ is called the ``context" of point $i$ by these
authors and was computed for recognition of 2-dimensional shapes via registration. Extending this concept to 3D objects is harder as higher dimensional ``histograms" need to be {computed} at each point to locally describe the points in a neighborhood of a given point. 

In contrast, the lower part of the spectrum of the LB operator provides a {\em global} description of the shape, with more details included in higher eigenvalue-eigenvector pairs. Interestingly, the geometric interpretation in expression (\ref{2H}) of the LB operator indicates a similarity with the point histograms of \cite{Belongie2002} in the sense that it generates a local characterization at each point on the manifold (the normal whose length relates to the mean curvature). Using the spectrum of the LB operator results in a very effective summary of this local information, providing a global description of the geometry (and even providing some topological information, as discussed in Section 2) of the part.
As discussed, the LB spectrum results from the solution of the Helmholtz equation, which relates to the spatial part of the heat equation. It seems possible to also look at the {\em temporal} solution of the heat equation to try to find additional {\em local} intrinsic descriptors of the geometry of an object, a matter for further work.

A result due to Huber (see e.g., \citealt{Buser}) proves how two Riemannian surfaces have the same sequence of eigenvalues of their LB operator if and only if they have the same sequence of lengths of their closed geodesics, thus, in principle, the geodesics, evidently intrinsic, could be used for detecting part to part differences in a SPC scheme. For our prototypical part, Figure \ref{figGeo} illustrates some geodesics between a point and some others in the part under consideration, as well as the geodesic distances between the point and all others on the mesh defining the part. Geodesics, being intrinsic to the surface, are simpler to compute than the spectrum, but in practice, the question, if one wishes to apply Huber's theorem, is that geodesics are greatly affected by noise \citep{lee2010spectral} and it is not clear how many geodesic distances are needed to compare between 2 objects, a manufactured/scanned one, and a noise-free CAD model, in order to find local defects on the surface of the manufactured part. In this section, we report our experiments with geodesic distances and other more robust {\em intrinsic} distance functions based on the heat kernel defined on the surface of a part.

\begin{figure}[h]
\begin{center}
\includegraphics[scale=0.3]{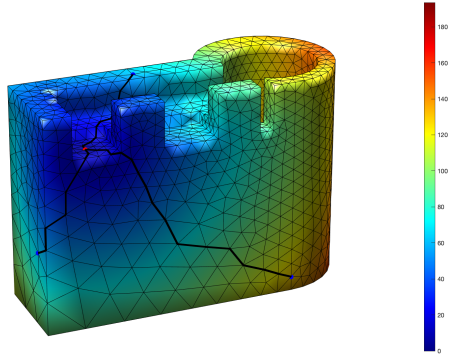}
\caption{Illustration of geodesics and geodesic distances on a mesh (triangulation) of a noise-free part: darker lines are the geodesics between a point and three other points on the triangulation (mesh) of our prototypical part. Color corresponds to geodesic distances from the first point over the triangulation, with redder colors indicating longer distances.}
\label{figGeo}
\end{center}
\end{figure}


Given  the spectrum $\{\lambda_i\}$ and associated eigenfunctions $\{\phi_i()\}$ of the LB operator on a general Riemannian manifold $\mc M$ (not necessarily flat euclidean space), the {\em heat kernel} is given by the decomposition:
\be
k_t(x,y) = \sum_{i=0}^{\infty} e^{-\lambda_i t} \phi_i(x) \phi_i(y) .
\label{heatKernel}
\ee
which, as mentioned before,  satisfies the heat equation. For small values of $t$,  $k_t(x,y)$ reflects {\em local properties} of $\mc M$ around point $x$; for larger values of $t$ it reflects the {\em global} structure of $\mc M$ from the point of view of $x$. It is, in a sense, also conveying analogous information as \cite{Belongie2002} ``shape contexts". There is a close relationship between the heat kernel and geodesic distances $g(x,y)$ for $x,y \in \mc M$:
\[
-g(x,y)^2 = \lim_{t \rightarrow 0} 4 t \log k_t(x,y)
\]
(see \citealt{SunHKS2009}), which can be shown {directly by} taking the limit to {$k_t(x,y) \approx \frac{1}{4 \pi t} e^{-g(x,y)^2/(4t)}$} which is an approximation of (\ref{heatKernel}) for small $t$.

The heat kernel can be also be interpreted as the transition density function of a Brownian motion on $\mc M$ \citep{SunHKS2009}, or alternatively, the estimated heat kernel, using a discrete LB approximation based on a mesh, can be related to the transition probabilities of a random walk process defined on the mesh, a property used in computer graphics by some authors (see, e.g., \citealt{sinha2014}) for shape similarity assessment.

 \cite{SunHKS2009} propose to use the ``autodiffusion" $k_t(x,x)$ at different points on a mesh as a {``}Heat Kernel Signature" used to identify shapes of objects. A useful notion related to the heat kernel is the {\em diffusion distance} between $x,y \in \mc M$ defined by the interplay of heat from $x$ to $y$ and from $y$ to $x$:
\be
d_t^2(x,y) = k_t(x,x)+k_t(y,y)-2k_t(x,y) = \sum_i e^{-\lambda_i t}(\phi_i(x)-\phi_i(y))^2
\label{Diffusion distance}
\ee
The diffusion distances are intrinsic and more robust to changes in the mesh as they are an ``average" distance over all possible paths between two points on the mesh, and are not only the shortest path as the geodesic distances are, a property that has received recent attention in the Statistics and Manifold Learning literature (e.g., see \citealt{lee2010spectral}). This property is shared by heat kernels. This might be a {\em dis}advantage for SPC: changes in the mesh may reflect true changes in the shape of the parts that need to be detected. Figure \ref{fig13} illustrates the evolution, for different values of $t$, of the heat kernels and diffusion distances from a point on the prototype part shown in earlier figures, given the LB spectrum of its mesh. These are the quantities that were computed in the histograms of the last two columns of Figure \ref{fig14}, whose quantile-quantile (Q-Q) plots appear in Figure \ref{fig15} (below).  {While the heat kernel can be easily understood  as the diffusion in time of heat from a point, the diffusion distances have no easy interpretation from their time evolution plot shown in the figure.}

Consider the three parts displayed on the leftmost column of Figure \ref{fig14} (measurements in mm.), where (as before) we have added isotropic $N(0, 0.05^2 {\bf I}_3)$ noise to the points of a CAD model defined as a triangulation on a (rather small) point cloud with 1680 points. The euclidean distances between all points for the in-control and defective parts are displayed in the second column; these histograms are invariant, but not intrinsic. The geodesic distances between all points, computed on the triangulation (using the ``Fast marching algorithm" on the triangulation, see \citealt{KimmelBook}) is displayed in the third column; this is both invariant and intrinsic. 
\begin{figure}
\centering
\includegraphics[scale=0.4]{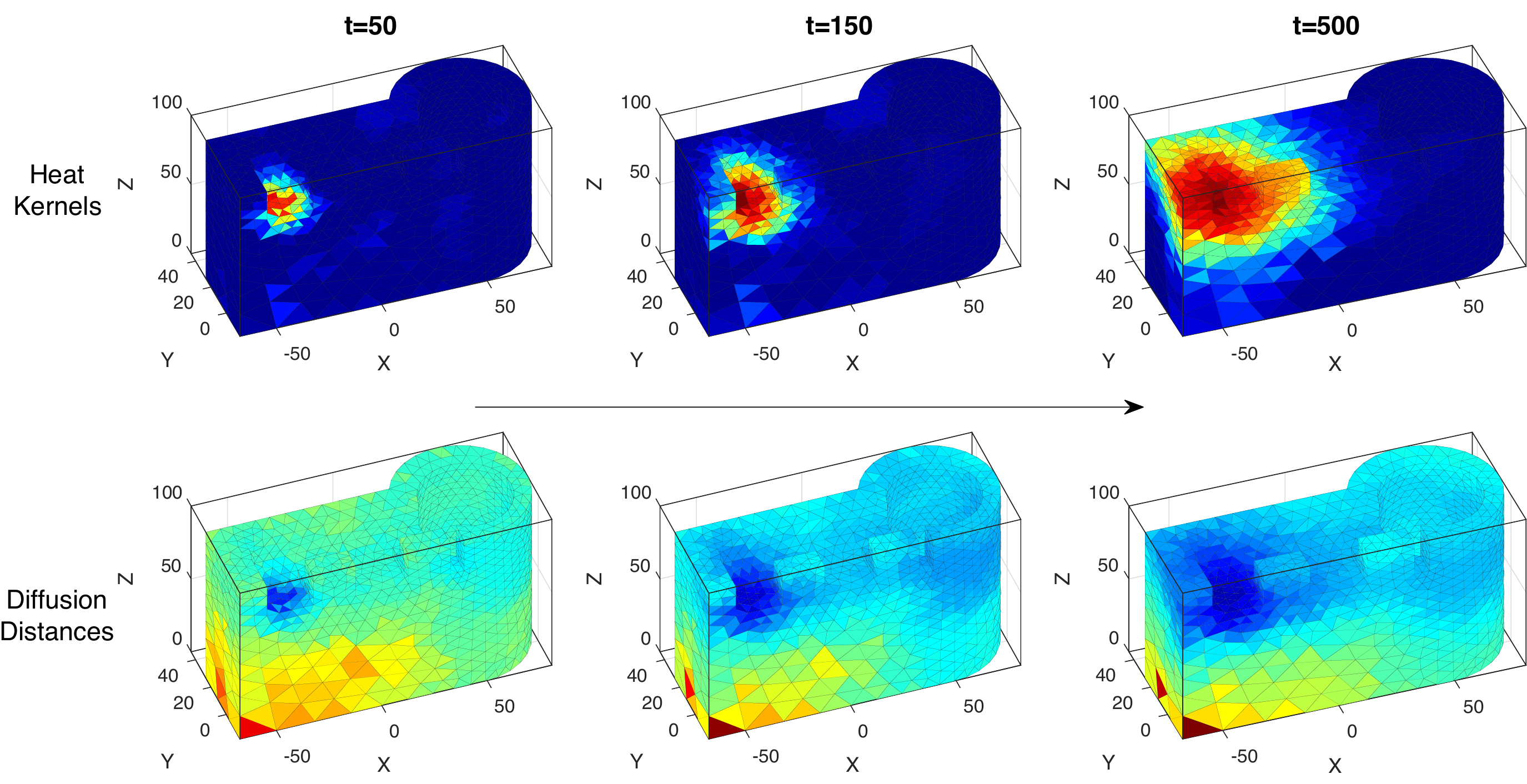}
\caption{The heat kernel (top) and the diffusion distances plotted from a point on the triangulation of a part as time increases from left to right. Redder colors indicate higher values. The heat kernel values tend to a steady-state value as $t$ diverges.}
\label{fig13}
\end{figure}

\begin{figure}[h]
\centering
\resizebox{14cm}{6cm}{\includegraphics{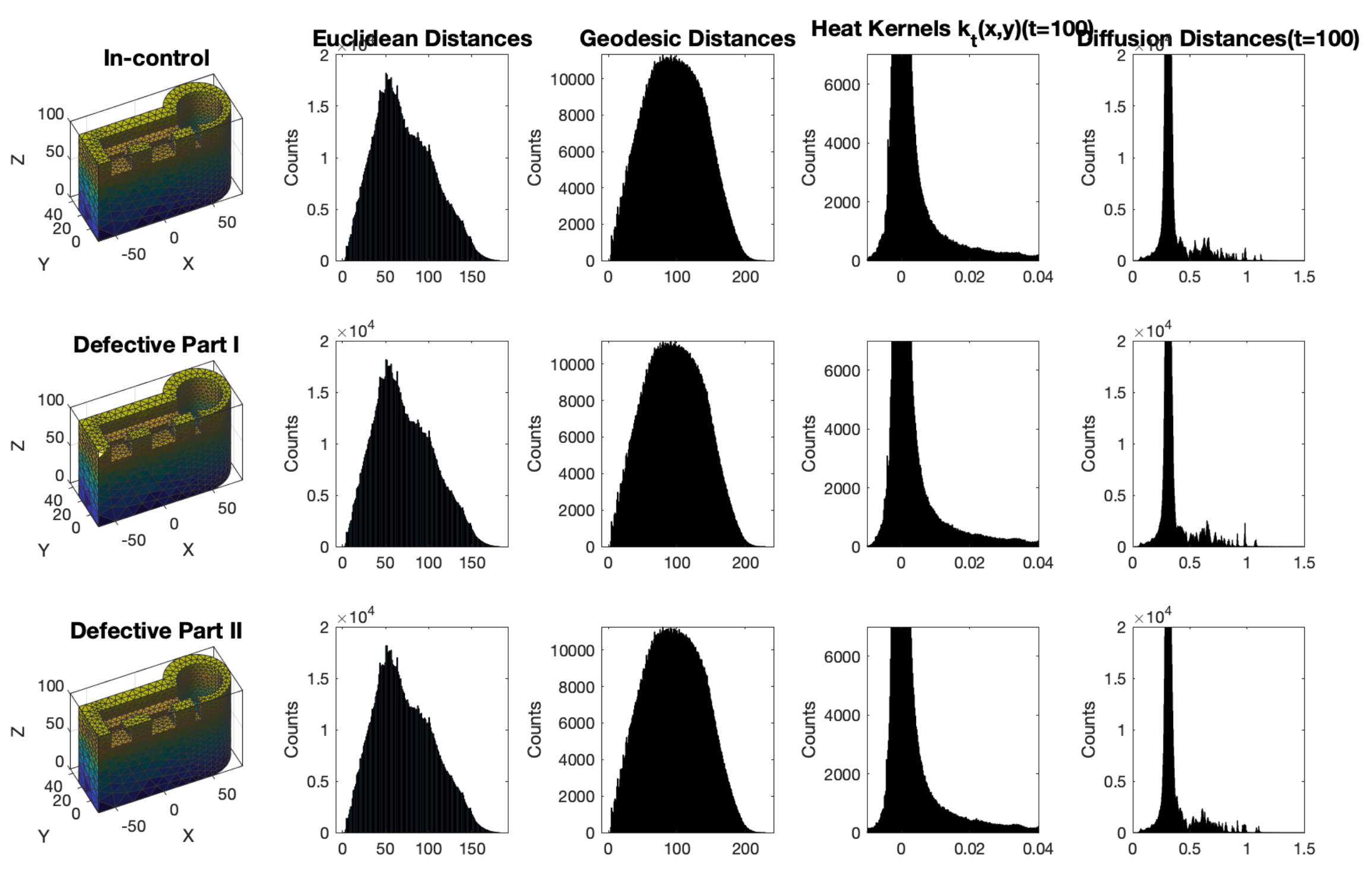}}
\caption{Histograms of the euclidean and geodesic distances, the heat kernels ($t=100$) and diffusion distances ($t=100$) for an in-control part (top row) and two types of parts with small local  defects, a chipped corner (middle row) and a protrusion on the top of the part (bottom row).  }
\label{fig14}
\end{figure}

Figure \ref{fig15} shows Q-Q plots for an in-control part and two of the types of local defects studied before, the chipped part \#1 shown in Figure \ref{fig14}, and the part with a protrusion on the edge at the top. Here the ``in control" distributions of the heat kernels and diffusion distances are plotted against the distributions of the defective parts, so the third {row} shows in-control vs. in-control distributions. We can see deviations from linearity in the profiles of both heat kernels and diffusion distances, but it is not clear what the role of the $t$ parameter in the ``detection" capabilities of a profile monitoring scheme would be. It seems plausible that smaller values of $t$ allow to detect {\em local} changes better, while larger values of $t$ would detect {\em global} changes better, but considerable more research is needed to assess this statement in order to provide useful tools for SPC. There are also deviations even in the third row, where an in-control part is plotted against another in-control part. Kolmogorov-Smirnov tests performed on all the cases in the figure are recognized as significantly different regardless of the type of the parts (in control or defective). This occurs due to the number of pairs of heat kernel or diffusion distances is so large (proportional to $n^2$, with $n$ being the number of points), that any slight difference is detected as significant.

\begin{figure}[h]
\centering
\resizebox{14cm}{6cm}{\includegraphics{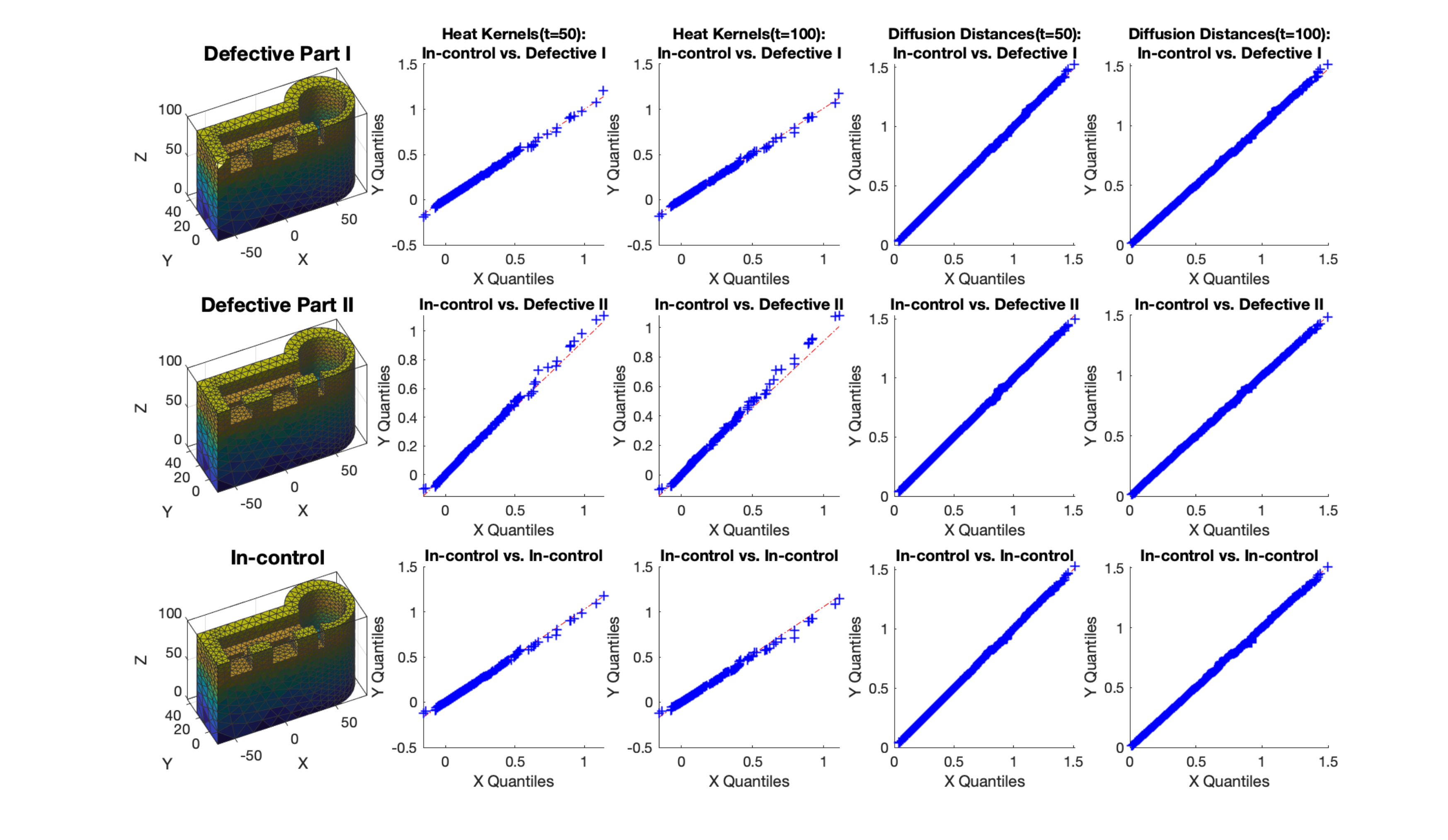}}
\caption{Q-Q plot of the heat kernels and diffusion distances between an in-control part and two types of parts with small local  defects, a chipped corner (top row) and a protrusion on the top of the part (bottom row). Here, $t=50$ (2nd and 4th column) and $t=100$ (3rd and 5th columns). }
\label{fig15}
\end{figure}

\subsection{Other spectral distances}

Both the heat kernel and the diffusion distances, and the other spectral distances we describe next, can be used as an alternative criterion for registering two parts, using the ``shape context" idea of \cite{Belongie2002} to find matches of points between two surfaces. The matches between points in a part that triggered an alarm and a CAD model could be then used as an additional post-alarm SPC diagnostic to determine where the produced part differs from the CAD model, analogously to the ICP post-alarm diagnostic presented earlier (this would require more computational effort than using ICP, as it requires both finding the spectrum and solving the combinatorial registration problem, but could provide better information specially when the defect is small and not evident to the eye). It is unclear, however, how to use these spectral distance diagnostic tools without recourse to registration or how to set in practice the $t$ parameter, present in both heat kernel and diffusion distances.

There are, however, spectral distances that eliminate the $t$ parameter. For a point $x$ on a surface $\mc M$, \cite{rustamov2007} defines the ``{G}lobal {P}oint {S}ignature" (GPS) as the map of $x$ into the infinite dimensional vector
\[
\mbox{GPS}({x}) = \left( \frac{1}{\sqrt{\lambda_1}} \phi_1(x), \frac{1}{\sqrt{\lambda_2}} \phi_2(x), \frac{1}{\sqrt{\lambda_3}} \phi_3(x),...\right)
\]
In practice, the GPS embedding is finite dimensional using the lower part of the discrete LB spectrum and the associated eigenvectors. The idea of embedding a point from a manifold $\mc M$ into a different dimensional space is a theme common to both Computer Graphics (e..g, ``Heat Kernel Signatures" \citealt{SunHKS2009}), where a {\em higher} dimensional embedding is sought to define a similarity metric in that space between points originally in $\mathbb{R}^3$, and Manifold Learning (see, e.g., \citealt{belkin2002} ``Laplacian eigenmaps"), where the goal is usually to find a {\em lower} dimensional embedding for understanding the data manifold structure. In the case of GPS, a distance function between two points on a 2-manifold can be defined by taking the inner product of their embeddings:
\be
G(x,y) = \langle \mbox{GPS}(x),\mbox{GPS}(y)\rangle = \sum_{i=1}^{\infty} \frac{\phi_i(x) \phi_i(y)}{\lambda_i}
\ee
which can be obtained also as the integral of the heat kernel $k_t(x,y)$ with respect to $t$. A distance (metric) between $x,y \in \mc M$ is then given by $\sqrt{G(x,y)}$. 

A distance (metric) related to the GPS distance is the so-called ``Commute {T}ime {D}istance", defined as:
\be
d_c(x,y)^2 = \sum_{i=1}^{\infty} \frac{(\phi_i(x)- \phi_i(y))^2}{\lambda_i}, \quad x,y \in \mc M
\ee
which is related to the GPS distance since $d_c(x,y)^2 = G(x,x) + G(y,y) - 2 G(x,y)$. \cite{lipman2010} indicate that both the GPS and the commute time distances have the disadvantage of diverging when $x=y$ since $\sum_{i=1}^{\infty} 1/\lambda_i \approx \sum_{i=1}^{\infty} 1/i$ diverges, and proposed instead  a modification, called ``biharmonic distance":
\be
d_b(x,y)^2 = \sum_{i=1}^{\infty} \frac{(\phi_i(x)- \phi_i(y))^2}{\lambda_i^2}, \quad x,y \in \mc M
\ee
which does not diverge when $x=y$ since $\sum_{i=1}^{\infty} 1/\lambda_i^2 \approx \sum_{i=1}^{\infty} 1/i^2 = \pi^2/6 < \infty$. In practice, all these distances need to be computed using the lower part of the spectrum of the discrete LB operator, up to a given eigenvalue-eigenvector pair index. The applicability of biharmonic distances to the SPC problem stems from the absence of a $t$ parameter, necessary in other spectral distances reviewed, and in principle they could be used for SPC by comparing their distribution with that of an in-control part using Q-Q charts as in Figure \ref{fig15}, but further work is necessary to examine this matter.

\end{appendix}

\end{document}